# Spatial proximity effects on the excitation of Sheath RF Voltages by evanescent Slow Waves in the Ion Cyclotron Range of Frequencies.

Laurent Colas[1,a], Ling-Feng Lu[1], Alena Křivská[2], Jonathan Jacquot[3], Julien Hillairet[1], Walid Helou[1], Marc Goniche[1], Stéphane Heuraux[4] and Eric Faudot[4].

[1]*CEA, IRFM, F-13108 Saint-Paul-Lez-Durance, France.*
[2]*LPP-ERM-KMS, TEC partner, Brussels, Belgium.*
[3]*Max-Planck-Institut für Plasmaphysik, Garching, Germany.*
[4]*Institut Jean Lamour, UMR 7198, CNRS-University of Lorraine, F-54506 Vandoeuvre Cedex, France.*

[a)]Corresponding author: laurent.colas@cea.fr

**Abstract.** We investigate theoretically how sheath radio-frequency (RF) oscillations relate to the spatial structure of the near RF parallel electric field $E_{//}$ emitted by Ion Cyclotron (IC) wave launchers. We use a simple model of Slow Wave (SW) evanescence coupled with Direct Current (DC) plasma biasing *via* sheath boundary conditions in a 3D parallelepiped filled with homogeneous cold magnetized plasma. Within a "wide sheaths" asymptotic regime, valid for large-amplitude near RF fields, the RF part of this simple RF+DC model becomes linear: the sheath oscillating voltage $V_{RF}$ at open field line boundaries can be re-expressed as a linear combination of individual contributions by every emitting point in the input field map. SW evanescence makes individual contributions all the larger as the wave emission point is located closer to the sheath walls. The decay of $|V_{RF}|$ with the emission point/sheath poloidal distance involves the transverse SW evanescence length and the radial protrusion depth of lateral boundaries. The decay of $|V_{RF}|$ with the emitter/sheath parallel distance is quantified as a function of the parallel SW evanescence length and the parallel connection length of open magnetic field lines. For realistic geometries and target SOL plasmas, poloidal decay occurs over a few centimeters. Typical parallel decay lengths for $|V_{RF}|$ are found smaller than IC antenna parallel extension. Oscillating sheath voltages at IC antenna side limiters are therefore mainly sensitive to $E_{//}$ emission by active or passive conducting elements near these limiters, as suggested by recent experimental observations. Parallel proximity effects could also explain why sheath oscillations persist with antisymmetric strap toroidal phasing, despite the parallel anti-symmetry of the radiated field map. They could finally justify current attempts at reducing the RF fields induced near antenna boxes to attenuate sheath oscillations in their vicinity.

## 1. CONTEXT AND MOTIVATIONS

In magnetic fusion devices, non-linear wave-plasma interactions in the Scrape-Off Layer (SOL) often set operational limits for Radio-Frequency (RF) heating systems *via* impurity production or excessive heat loads[Noterdaeme1993]. Peripheral Ion Cyclotron (IC) power losses are generally attributed to RF sheath rectification. How this non-linear process depends on the geometry and electrical settings of the IC wave launchers remains largely unknown, despite crucial technological implications. In low-frequency small capacitive plasma discharges, sheath rectification has been successfully modelled in analogy with a double Langmuir probe driven by an oscillating voltage $\tilde{V}$[Chabert2011]. In the absence of more elaborate theory in realistic tokamak geometry over large scale lengths, this simple formalism was also widely applied near IC antennas, without strong justification (e.g. in [Perkins1989]). Along this line of thought, the RF field parallel to the confinement magnetic field $\mathbf{B_0}$, integrated along isolated open magnetic field lines, $\tilde{V}=|\int E_{//}.dl|$, has often been used as a quantitative indicator of local RF sheath intensity in the vicinity of IC antennas, e.g. in [D'Ippolito1998],[Colas2005], [Mendes2010], [Garrett2012], [Milanesio2013], [Qin2013], [Campergue2014]. In this exercise one often uses $E_{//}$ fields from full-wave linear electromagnetic simulations where the plasma is in direct contact with metallic walls (i.e. without sheaths) [Milanesio2009][Jacquot2015].

In tokamak experiments, qualitative correlation was noticed between the evolution of $\tilde{V}=|\int E_{//}.dl|$ and that of heat load intensity [Colas2009],[Campergue2014] or plasma radiation [Qin2013] [Colas2009]. Yet recent tokamak measurements challenge the relevance of $\tilde{V}$ as an indicator of RF sheath intensity. For example the line integral is expected to vanish in presence of a RF field map anti-symmetric along the parallel direction. This is nearly the case with anti-symmetric toroidal phasing of the IC poloidal strap arrays. Although the wave-plasma peripheral interaction observed experimentally is weaker with two straps phased [0,π] than with [0,0] phasing[Colas2009], [Bobkov2015], it is not suppressed. Similar experimental results were obtained with more straps [Lerche2009],[Jacquet2011], [Jacquet2013], [Wukitch2013]. The magnetic field pitch with respect to the toroidal direction is often invoked to interpret the persistence of RF sheaths, together with antenna boxes breaking the anti-symmetry [Colas2005]. On ASDEX-Upgrade, closing the box corners with metallic triangles did not suppress the local impurity production[Bobkov2010]. To mitigate the effect of magnetic field pitch, a field-aligned antenna was designed for ALCATOR C-mod. In comparison with a toroidally-aligned antenna, it was predicted to reduce $|\tilde{V}|$ on open flux tubes with large toroidal extension on either sides of the IC wave launcher("long field lines") [Garrett2012]. The expected reduction was significant with [0000] phasing of the 4-strap array. Experimental comparisons on ALCATOR C-mod revealed a reduced Molybdenum contamination when using the field-aligned antenna [Wukitch2013]. But the plasma potentialmeasured on magnetic field lines connected to the antenna hardly varied, andthe wave-SOL interaction was not suppressed with [0000] strap phasing.

At CEA, a prototype Faraday Screen (FS) was designed to reduce $|\tilde{V}|$ over "long field lines", by interrupting all parallel RF current paths on its front face [Mendes2010]. When compared to an antenna equipped with standard FS on Tore Supra (TS), the new FS exhibited similar heat load spatial distribution, but the measured RF wave-SOL interaction was more intense and more extended radially [Colas2013]. In a series of TS experiments, the left-right ratio ofstrap voltage amplitudes was varied. Over this scan, the antenna side limiter near the strap with higher voltage heated up, while the remote limiter cooled down. A similar toroidal unbalanceon ASDEX Upgradeproduced opposite variations of RF currents amplitudesmeasured at the surface of two opposite antenna limiters [Bobkov2015]. In this experiment with [0,π] phasing, in order to minimize the collected RF current, the RF voltage imposed on the remote strap was approximately twice higher than the voltage on the strap near the side limiter. Thesetrendscan hardly be explained using a single physical parameter simultaneously relevant at both extremities of the same openmagnetic field line, $\tilde{V}$ or any other one. Besides, in $\tilde{V}$, all the points along the integration path play the same role. Theexperimental observationsrather suggest that the toroidal distance between radiating elements andthe observed walls might play a role in the RF-sheath excitation.From this paradigm,minimizing the local RF electric field amplitudes near the antenna limiters, still evaluated in the absence of RF sheaths, was proposed to mitigate RF-sheath generation on new ASDEX-Upgrade antennas[Bobkov2015]. This alternative heuristic procedure also deserves justification from first principles.

The "double probe" analogyimplicitly assumes that each open magnetic field line behaves as electrically isolated from its neighbors. This is questionable in highly conductive plasmas, although the conductivity is far larger along **B₀** than transverse to it. Collected RF currents on ASDEX Upgrade also challenge this picture. Early attempts at improving the "double-probe" models suggest that the exchange ofcurrents between neighboring flux tubes decouplesthe sheaths at the two extremities of the same open field line[Rozhansky1998],[NGadjeu2011], [Faudot2013], [Jacquot2011]. The self-consistent spatio-temporal description of RF electric fields and RF currents, i.e. electrodynamics, has been long developed in the context of IC antennas radiating in magnetized plasmas, but in the absence of sheaths[Milanesio2009],[Jacquot2015],[Lu2016a]. The RF plasma conductivity is then incorporated in a time-dispersive dielectric tensor [Stix1992]. Unlike capacitive RF discharges, tokamak field mapsfeature spatially inhomogeneous RF electric fields in the quasi-neutral plasma surrounding the wave launchers[Colas2005], [Mendes2010], [Bobkov2010], [Garret2012]. The distances between radiating elements and observation points, parallel and transverse to $B_0$, can then play a role *via* the propagation of RF waves.

These experimental and theoretical considerations motivated several models coupling RF wave propagation and Direct Current (DC) SOL biasing *via* RF and DC sheath boundary conditions applied at the plasma-wall interfaces [D'Ippolito2006], [Myra2008], [D'Ippolito2009], [D'Ippolito2010],[Myra2010], [Kohno2012], [Kohno2012a], [Colas2012], [Jacquot2014], [Jenkins2015], [Kohno2015]. Within this general framework, the double probe analogy was assessed and spatial proximity effects were investigated. Simple situations were exhibited where sheath oscillations exist on open magnetic field lines for which $\tilde{V}=0$ [D'Ippolito2006].Other situations were studied for which $\tilde{V}$over-estimates the sheath voltage[D'Ippolito2009].In presence of a localized source of evanescent $E_{//}$, situated half-way between the two field line extremities, it was noticed that the sheath excitation was reduced when the connection length exceeded the ion skin depth [Myra2010]. Inthese simple situationsthe two extremities of the same open magnetic field lines generally behaved similarly, for symmetry reasons. Realistic RF-sheath simulations of the Tore Supra antenna environment, limited to the Slow Wave (SW),explored toroidally asymmetric RF field maps and reproduced qualitatively the observed left-right asymmetric heat loads and other experimental measurements[Jacquot2014]. Similar efforts are underway to interpret the ASDEX-Upgrade phenomenology[Křivská2015] [Jacquot2015].In simulations of the ITER antenna, parallel proximity effects were evidenced numerically, but not interpreted[Colas2014].

This paper reformulates one of the above-mentioned models of coupled RF wave propagation and DC SOL biasing, called Self-consistent Sheath and Waves for Ion Cyclotron Heating – Slow Wave (SSWICH-SW)[Colas2012] [Jacquot2014].Within restrictive assumptions on simulation domain shape, radial profiles, wave amplitude and polarization, weexplain andquantify spatial proximity effects and left-rightsheath asymmetries.Calculus is easier in a "wide sheath" regime, for which the SW propagation and subsequent excitation of sheath RF oscillations becomes a linear problem. Within this asymptotic limit, valid for intense DC biasing, the amplitude of the sheath oscillating voltages can be re-expressed as a *weighted* integral of $E_{//}$. This offers an alternative to $\tilde{V}$ for assessing RF-sheath excitation, with stronger theoretical justification. In the integral, proximity effects arise from the spatial dependence of the weight function (a Green's function for the linear problem).The SW evanescence between its emission point and the sheath spatial location appears to strongly affect this spatial dependence. Other parametric dependences are also evidenced. After briefly recalling the SSWICH-SWmodel, the paper investigates the Green's functions in two and three dimensions for a parallelepiped simulation domain filled with homogeneous plasma and **B₀** normal to the lateral walls.The geometricalproperties of the Green's functions are quantified using characteristic scale-lengths of the problem. In light of thisre-formulated model we finally re-interpret the experimental observations summarized above. Concrete implications of the results arediscussed, as well as some limitations of the approach.

## 2.  COUPLING SLOW WAVE PROPAGATION AND DC PLASMA BIASING BY RADIO-FREQUENCY (RF) SHEATHS

### 2.1 Outline of SSWICH-SW model

Our minimal model of coupled RF wave propagation and DC plasma biasing,SSWICH-SW,was detailed in references [Colas2012],[Jacquot2014]and is briefly summarized here. The simulation domain, sketchedon Figure 1,features a collection of straight open magnetic flux tubes in a slab idealization of a tokamak SOL plasma. Two versions of the geometry will be used: a three dimensional (3D) model with boundaries parallel to the poloidal direction (*y*); as well as a 2D cut into the above 3D model along the radial direction (*x*) and parallel to the confinement magnetic field **B₀** (direction *z*). Both simulation domains are filled with cold magnetized plasma homogeneous along direction *y,* with possibly radial variation. Inner and outer boundaries of the domain are normal to *x*, while material boundaries of the fusion device are either parallel or normal to **B₀**. Thisallows versatile geometries with radial profiles of the plasma parameters and private SOLs, sketched as gray levels, as well as protruding material objects, e.g. IC antenna side limiters(see [Jacquot2014]), intercepting the magnetic field lines and developing sheaths.

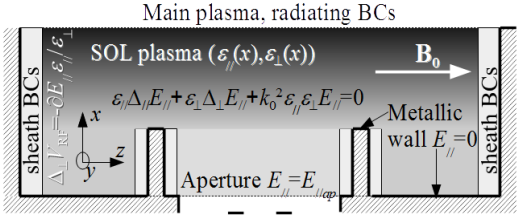

**FIGURE 1**: 2D (radial/parallel) cut into SSWICH general 3D simulation domain (not to scale). Main equations and notations used in the paper. The gray levels are indicative of the local plasma density. Light gray rectangles on boundaries normal to $\mathbf{B_0}$ feature the presence of sheaths, treated as boundary conditions in our formalism.

In this domain, the simulation process couples 3 steps self-consistently.

**Step 1: Slow Wave propagation**. The physical system is excited by a 2D (toroidal, poloidal) map of the complex parallel RF electric field $E_{//\text{ap}}(\mathbf{r_0})$, radiated by an IC antenna and prescribed at an aperture in the outer radial boundary of the simulation domain. The input RF field map presently needs to be computed *a-priori*, generally without sheaths, by an external antenna code. No restriction is imposed *a-priori* on the spatial distribution $E_{//\text{ap}}(\mathbf{r_0})$ that carries the information about antenna geometry and its electrical settings. For our particular purpose in this paper it is necessary to compare several peaked distribution of fields with the same $\int E_{//\text{ap}}.dl$ but different parallel locations of the field maximum, at fixed connection length $L_{//}$. We would also like to assess how far the RF field at a given poloidal position affects the sheaths at other poloidal positions. We would finally like to investigate RF field maps that are asymmetric with respect to the middle of the open magnetic field lines. These classes of field maps were scarcely considered in earlier literature [D'Ippolito2006], [Myra2010].

From the aperture, a time-harmonic cold Slow magneto-sonic Wave (SW) with pulsation $\omega_0$ propagates across the whole simulation domain according to equation [Stix1992]

$$\varepsilon_\perp \Delta_\perp E_{//} + \varepsilon_{//} \Delta_{//} E_{//} + k_0^2 \varepsilon_{//} \varepsilon_\perp E_{//} = 0 \qquad (1)$$

With $\Delta_{//}=\partial_{zz}^2$ the parallel Laplace operator, $k_0=\omega_0/c$ the vacuum wavenumber, and $(\varepsilon_{//},\varepsilon_\perp)$ the diagonal elements of the local cold plasma dielectric tensor [Stix1992].

In 3D $\Delta_\perp=\partial_{xx}^2+\partial_{yy}^2$, while in 2D $\Delta_\perp=\partial_{xx}^2-k_y^2$, where $k_y$ is a wavevector in the ignorable (poloidal) direction $y$. These transverse derivatives couple adjacent magnetic field lines, unlike the simplest "double probe" models. Equation (1) is subject to radiating conditions at the inner radial boundary, metallic conditions $E_{//}=0$ on material boundaries parallel to $\mathbf{B_0}$, and RF sheath boundary conditions (RF SBCs) at the parallel boundaries (see Figure 1). RF SBCs, first proposed in reference [D'Ippolito2006], will be further discussed.

**Step 2: RF oscillations of the sheath voltage.** When reaching the extremities of the open magnetic field lines, the SW fields $E_{//}$ generate oscillations $V_{RF}$ of the sheath voltage at the RF pulsation $\omega_0$. $V_{RF}$ is a complex quantity incorporating amplitude and phase information. The definition $\mathbf{E}_\perp = \pm\nabla_\perp V_{RF}$ at the sheath/plasma interface, combined with the relation $\nabla\cdot(\varepsilon\mathbf{E})=0$ valid all over the plasma, using $\mathbf{rot}_\perp\mathbf{E}_\perp=\mathbf{0}$ for the SW, yield a diffusion equation for the sheath oscillating voltages $V_{RF}$ along the boundaries normal to $\mathbf{B_0}$, including a source term due to the SW[Colas2012],

$$\varepsilon_\perp \Delta_\perp V_{RF}(x,y,z_{wall}) = \mp\varepsilon_{//}\partial_{//}E_{//}(x,y,z_{wall})$$
$$V_{RF}(x,y,z_{wall}) = 0 \text{ at boundary extremities} \quad (2)$$

Since the quantity $V_{RF}$ is only meaningful at sheaths, equation (2) applies only at the domain boundaries normal to $\mathbf{B_0}$ (see figure(1)).

**Step 3: Rectification of the sheath oscillations.** Due to the non-linear I-V characteristics of the sheath, the RF oscillations of the sheath voltage are rectified into enhanced DC biasing of the SOL plasma. Several DC biasing models exist in the literature[D'Ippolito2006], [Myra2015]. These will not be detailed here, but the DC plasma potential $V_{DC}$ is an increasing function of the RF voltage amplitudes $|V_{RF}|$. The DC voltage drop across the sheaths affects their width *via* the Child Langmuir law, and consequently their RF admittance and the RF SBCs applied for $E_{//}$[D'Ippolito2006]. Therefore all steps defined above generally need to be iterated till convergence is reached[Jacquot2014]. However for sheaths wider than a characteristic value, the RF SBCs were found nearly independent of the sheath widths[Colas2012][Kohno2012]. For $\mathbf{B_0}$ normal to the wall the asymptotic RF SBCs simplify into $E_{//}=0$. This wide sheath limit was used as a first guess to start the iterative resolution of the model. In realistic Tore Supra simulations with self-consistent sheath widths, the near RF fields were intense enough to approach this "wide sheath" asymptotic regime[Jacquot2014].

## 2.2 Green's function reformulation of RF-sheath excitation with prescribed sheath widths

Step 3 is intrinsically non-linear, making the whole model non-linear when steps 1-3 are coupled self-consistently, as e.g. in[Myra2008], [Myra2010], [D'Ippolito2008], [D'Ippolito2009], [D'ippolito2010], [Jacquot2014]. However when sheath widths are prescribed in a non-self-consistent way in every point, steps 1-3 can be run successively. In particular this exercise can be done using the self-consistent spatial distribution of sheath widths once it is known. Equations (1) and (2) arelinear, together with their BCs. This property can be exploited to evidence and quantify spatial proximity effects in the RF sheath excitation.The superposition principle indeed allows re-expressing $V_{RF}(\mathbf{r})$ evaluated at any sheath boundary point **r**as the linear combination of contributions by every emitting point at position $\mathbf{r_0}$in the input RF field map.

$$V_{RF}(\mathbf{r}) = \int_{aperture} G(\mathbf{r},\mathbf{r_0}) E_{//ap}(\mathbf{r_0}) d\mathbf{r_0} \quad (3)$$

Relation (3) formally looks like the integral $\tilde{V}=\int E_{//}.dl$ used in the "double probe" model, with major differences however. 1°) $V_{RF}(\mathbf{r})$ relates to one sheath, whereas $\tilde{V}$ was applied between two electrodes. Depending on the parallel symmetry of the input RF field map, the two extremities of the same open field line can now oscillate differently. 2°) Rather than along each open field line, integration is now performed over the aperture, either in 1D or 2D depending on the considered geometry. Neighboring open field lines can therefore be coupled. 3°) A weighting factor $G(\mathbf{r},\mathbf{r_0})$is applied to $E_{//ap}(\mathbf{r_0})$, depending on the parallel and transverse distances from the field emission point $\mathbf{r_0}$to the observation point **r**at the sheath walls.

$G(\mathbf{r},\mathbf{r_0})$ is the solution of equations (1) and (2) with elementary excitation$E_{//ap}(\mathbf{r})=\delta(\mathbf{r}-\mathbf{r_0})$, i.e. a Green's function of the linear problem with one point source switched on in the input field map.$G(\mathbf{r},\mathbf{r_0})$ only carries information on the geometry of the simulation domain and on the SOL plasma parameters, while the input field map $E_{//ap}(\mathbf{r_0})$ accounts for the antenna properties. $V_{RF}(\mathbf{r})$ combines the two characteristics. In a rectangular box filled with homogeneous plasma, the Dirac source term can be decomposed into eigenmodes of wave propagation in the box with sinusoidal variation in the parallel and poloidal directions. This can be done either in the wide-sheath limit [Colas2012] or when sheath widths are assumed uniform all over the box [Myra2010]. In these simple casesa formal Fourier correspondence exists between the Green's function approach and these earlier spectral methods.

The formal simplicity of relation (3) hides two main difficulties.

-The initial non-linear problem is apparently turned into a linear relation. But computing the self-consistent sheath widths requires solving the fully coupled problem that is non-linear. Howeverin the wide sheaths limit the RF electric fields can be computed without knowing *a-priori* the sheath widths spatial distribution[Colas2012]. Below we will work within this limit. This imposes restrictions on the wave amplitude, but not on its spatial distribution, the main topic of this paper. Unlike [Myra2010] we will therefore not attempt at self-consistency: the resulting $V_{RF}$ is the one obtained after only one turn around our iterative simulation loop.

-While the $V_{RF}$ re-formulation applies for complex geometries, in presence of radial density gradients and any spatial distribution of the prescribed sheath widths, the Green's functions are hard to obtain in these very general cases. Consequently this approach is generally less efficient than alternative ways to calculate oscillating voltages, e.g. spectral methods [Myra2010], [Colas2012] or finite elements [Kohno2012] [Jacquot2014]. Its main merit is to characterize explicitly the relation of $V_{RF}(\mathbf{r})$ to the spatial structure of the SW field, our particular purpose. In order to get insight into these properties we treat below simple cases in parallelepiped geometry that are tractable semi-analytically.

## 3. PROXIMITY EFFECTSON THE EXCITATION OF SHEATH RFVOLTAGES BY EVANESCENT SLOW WAVES IN 2D

We restrict firstour initial geometry to a 2-dimensional (2D) rectangular domain of dimensions ($L_{//}$, $L_{\perp}$) in the (parallel, radial) directions, filled with cold magnetized plasma homogeneous in all directions. In the ignorable direction $y$, spatial oscillations as $\exp(ik_y y)$ are assumed for RF quantities. The geometry is summarized in figure 2. The simulation domain is representative of the private SOL in front of an ICRF wave launcher, with $L_{\perp}$ the radial protrusion of (simplified !) antenna side limiters and $L_{//}$ the parallel distance between their internal faces. $E_{//}$ is prescribed at antenna aperture plane $x=0$. Radiating boundary conditions for $E_{//}$ are enforced at the inner boundary $x=L_{\perp}$, andasymptotic RF sheath BCs at parallel extremities $z=\pm L_{//}/2$. This simplified geometry shares some similarity with the situation treated in [Myra2010]. However in this earlier publication $k_y=0$ was assumed and the simulated domain was unbounded in the radial direction ($L_{\perp}\to\infty$). In their detailed calculations only one particular class of input field maps was considered: Gaussian peaks whose top was always located half-way between the two sheath walls. For symmetry reasons, with this class of field maps the sheaths at both open field line extremities could be characterized by one single voltage. Focus was put on obtaining self-consistent sheath voltages, under the following assumptions:

-Self-consistent sheath widths were assumed the same at the two extremities of the same open magnetic field line

-Sheath widths (and hence DC plasma potentials) did no vary in the radial direction.

-On average over many RF periods, sheaths were assumed to float in every point.

In our simplified problem relation (3) becomes

$$V_{RF}(x, z = \pm L_{//}/2) = \pm \int_{-L_{//}/2}^{L_{//}/2} E_{//ap}(z_0) G_{2D}(x, k_y, \mp z_0) dz_0 \qquad (4)$$

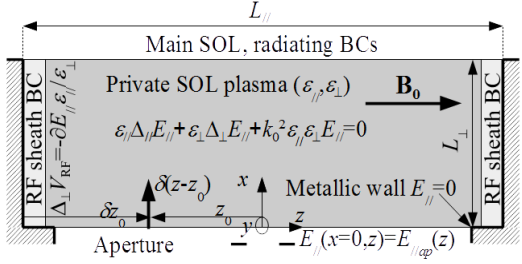

**FIGURE 2**: Generic 2D simulation domain (not to scale). Main equations and notations used in parts 3 and 4. $x=0$ at aperture. Light gray rectangles on boundaries normal to $\mathbf{B}_0$ feature the presence of sheath boundary conditions.

Green's function $G_{2D}(x,z_0)$ is non-dimensional and can be obtained from equations (1) and (2) at the left boundary with excitation $E_{//}(x=0,z)=E_{//\mathrm{ap}}(z)=\delta(z-z_0)$ (see Figure 2). The sheath properties at the right boundary can be deduced by changing appropriate signs in (4). $V_{RF}$ from relation (4) have the same magnitude at the two extremities of the same open magnetic field line only if the input field map is symmetric or anti-symmetric along $\mathbf{B}_0$.

### 3.1 Characteristic scale-lengths

Introducing characteristic squared lengths

$$L_x^2 = \left[ k_y^2 - k_0^2 \varepsilon_{//} \right]^{-1} \quad ; \quad L_z^2 = \left[ \varepsilon_\perp \left( k_y^2 / \varepsilon_{//} - k_0^2 \right) \right]^{-1} = L_x^2 \varepsilon_{//} / \varepsilon_\perp \qquad (5)$$

equation (1) can be recast into a standard form in the normalized space coordinates $X=x/|L_x|$ and $Z=z/|L_z|$

$$s_x \partial_{XX}^2 E_{//} + s_z \partial_{ZZ}^2 E_{//} - s_x s_z E_{//} = 0 \qquad (6)$$

Where $s_x$ (resp. $s_z$) are the signs of $L_x^2$ (resp. $L_z^2$). Four cases need to be distinguished, corresponding to the four combinations of signs. If both signs are the same, equation (6) is elliptic and describes propagative (negative signs) or evanescent waves (positive signs) qualitatively similar to those in ordinary dielectric materials (i.e. the anisotropy of the magnetized plasma is a matter of length stretching). If signs are opposite equation (6) becomes hyperbolic and describes propagating waves with resonant cone properties comparable to Lower Hybrid waves in tokamaks [Stix1992]. In practice, $s_x<0$ corresponds to unrealistically low densities for IC waves in the SOL of tokamaks, for which sheaths have limited operational consequences. For $s_x>0$, $s_z$ is the sign of $\varepsilon_\perp$ and could possibly change over the SOL: $s_z<0$ prevails in a tenuous plasma that might exist in an IC antenna box... $s_z>0$ corresponds to typical plasma parameters measured in the SOL surrounding IC antennas on Tore Supra [Jacquot2014] and ASDEX-Upgrade[Křivská2015]. Below we study specifically this latter case and treat $L_x$ and $L_z$ as real positive quantities. $\varepsilon_\perp=0$ corresponds to the Lower Hybrid resonance and is associated to very large $L_z$. Equation (1) implicitly assumes a scale separation between the SW and the Fast Wave. Close to the Lower Hybrid resonance this separation needs to be revisited to allow a possible mode conversion between the two wave polarizations. This is however outside the scope of the present paper.

Lateral boundaries at finite distance from the emission points also introduce $L_{//}$ as a characteristic length of the wave propagation model. Excitation finally deserves normalization

$$E_{//\text{ap}}(z)=\delta(z-z_0)=L_z^{-1}\delta(Z-Z_0) \qquad (7)$$

A similar dimensional analysis can be made for equation(2) at the left boundary, using the normalized coordinates from (5)

$$\partial_{XX}^2 G_{2D} - k_y^2 L_x^2 G_{2D} = \partial_Z L_z E_{//} \qquad (8)$$

where from (7) $L_z E_{//}$ is non-dimensional. Equation(8) introduces the extra scale-length $k_y^{-1}$ into the problem, *via* the dimension-less parameter $k_y^2 L_x^2 = [1-k_0^2 \varepsilon_{//}/k_y^2]^{-1}$. Besides, the boundary conditions involve $L_\perp$.

In principle, all the geometrical properties of $G_{2D}(x,k_y,z_0)$ can be expressed in terms of $(x, z_0)$ and the characteristic lengths. Throughout the paper typical examples will illustrate our calculations with realistic geometrical, plasma and RF parameters used for ASDEX-Upgrade simulations in[Křivská2015]. Dielectric properties correspond to a standard D[H] minority heating scheme at frequency $f_0$=30MHz, with local magnetic field $B_0$=1.44T and L-mode SOL density $n_e$=8.3×10$^{17}$m$^{-3}$ in the antenna region. Geometry refers to ASDEX-Upgrade 2-strap antennas. Simulation parameters are $\varepsilon_{//}$=-74659, $\varepsilon_\perp$=-24.31, $k_0$=0.63m$^{-1}$, $L_{//}$=0.66m, $L_\perp$=12mm. For this particular case $L_x$=5.8248mm while $L_z$=0.3228m for $k_y$=0. For this realistic example, the parallel evanescence length is thus half the parallel extension of the antenna, while the transverse evanescence length is a small fraction of the poloidal height for the antenna.

### 3.2 2D electric field maps

The solution to equation (6) can be built from well-known results for the 2D Helmholtz equation in isotropic cylindrical geometry using modified Bessel functions of the second kind $K_j$ (*j* integer [Angot1972]). The method of images [MF1953] is then applied to account for the parallel boundary conditions at finite distance from the emitting point. Using the normalized coordinates $(X,Z)=(x/L_x, z/L_z)$ the field map writes

$$E_{2D}(x,z,z_0) = L_z^{-1} \sum_{n=-\infty}^{+\infty} (-1)^n F_{2D}(X, Z-Z_n) \quad ; \quad Z_n = [nL_{//} + (-1)^n z_0]/L_z \quad (9)$$

Where

$$F_{2D}(X,Z) = \frac{X}{\pi R} K_1(R) \quad ; \quad R^2 = X^2 + Z^2 \quad (10)$$

Here argument $R$ is the (normalized) distance to the emitting source. $F_{2D}$ describes the SW evanescence from a boundary point source in $(X,Z)=(0,0)$, in absence of parallel boundaries. For a fixed $X$ and $Z>>X>1$, $F_{2D}$ decays as $\sim\exp(-Z)$ along the parallel direction. $F_{2D}$ is null in $X=0$, except in $Z=0$ where the source term creates a singularity.

At the left boundary the RF sheath voltage excitation in (2) depends on the parallel derivative $\partial_z E(x, z=-L_{//}/2, z_0)$, with

$$\partial_z E_{2D}(x,z,z_0) = L_z^{-2} \sum_{n=-\infty}^{+\infty} (-1)^n \partial_Z F_{2D}(X, Z-Z_n)$$

$$\partial_Z F_{2D}(X,Z) = -\frac{XZ}{\pi R^2} K_2(R) \quad (11)$$

$z_0=+L_{//}/2$ corresponds to a source point near the right parallel boundary of the simulation domain. When $z_0=+L_{//}/2$ and $z=-L_{//}/2$, $Z_{2p}=Z_{2p+1}$, for all $p$ integer in summation (9) whence

$$\partial_z E_{2D}(x, -L_{//}/2, +L_{//}/2) = 0 \quad (12)$$

When the source point gets very close to the left sheath wall it is convenient to introduce $\delta Z_0=(z_0+L_{//}/2)/L_z$, the normalized parallel distance from the source point $z=z_0$ to the left boundary $z=-L_{//}/2$ (see figure 2). For sufficiently small $\delta Z_0$, $n=0$ and $n=-1$ become the dominant terms in the summation (9)

$$\partial_z E_{2D}(x, z=-L_{//}/2, z_0) \approx 2L_z^{-2} \partial_Z F_{2D}(X, \delta Z_0) = \frac{2X\delta Z_0}{\pi R_0^2 L_z^2} K_2(R_0) \quad (13)$$

Formula (13) shows that $\partial_z E_{2D}(x,z=-L_{//}/2,z_0)$ tends to 0, except perhaps in $x=0$, where $R_0$ vanishes. In the limit $R<<1$, $K_2(R) \sim 2/R^2$ [Angot1972] and

$$\partial_z E_{2D}(x, z=-L_{//}/2, z_0) \approx \frac{4X\delta Z_0}{\pi L_z^2 [X^2 + \delta Z_0^2]^2} \quad (14)$$

To shed light into the limit behavior $\delta Z_0 \to 0$, $X \to 0$ let us integrate with respect to $X$

$$\int_X^\infty \partial_z E_{2D}(x, z = -L_{//}/2, z_0) dX \approx \frac{2\delta Z_0}{\pi L_z^2 [X^2 + \delta Z_0^2]} \quad (15)$$

Integrating once again yields

$$\int_0^X \frac{2\delta Z_0}{\pi L_z^2 [X'^2 + \delta Z_0^2]} dX' = \frac{2}{\pi L_z^2} \arctan\left(\frac{X}{\delta Z_0}\right) \xrightarrow{\delta Z_0 \to 0} \frac{1}{L_z^2} \text{ whetever} X \quad (16)$$

Whence in the limit $\delta z_0 \to 0$, $x \to 0$

$$\partial_z E_{2D}(x, z = -L_{//}/2, z_0) \approx \frac{2}{L_z^2} \partial_X \delta(X) = 2\frac{\varepsilon_\perp}{\varepsilon_{//}} \partial_x \delta(x) \quad (17)$$

The limit $\delta Z_0 >> 1$, $Z_1 >> Z_0$ is accessible if $L_{//} >> L_z$. $Z_1 >> Z_0$ implies that $n=0$ and $n=-1$ are still the dominant terms in the summation (9), so that formula (15) applies. In the limit of large arguments $K_2(R) \sim [\pi/2R]^{1/2} \exp(-R)$ [Angot1972], so that

$$\partial_z E_{2D}(x, z = -L_{//}/2, z_0) \approx -\sqrt{\frac{2}{\pi}} \frac{X \delta Z_0}{L_z^2 [X^2 + \delta Z_0^2]^{5/2}} \exp\left[-\left(X^2 + \delta Z_0^2\right)^{1/2}\right] (18)$$

If $\delta Z_0 >> X > 1$, then $\partial_z E_{2D}(x, z=-L_{//}/2, z_0)$ decreases as $\sim \exp(-\delta z_0/L_z)$ as the source point moves away from the sheath wall. The characteristic length $L_{//}$ does not appear explicitly in expression(18). Indeed this length is related to the boundary conditions.

**3.3 Geometrical properties of 2D Green's function for the sheath oscillating voltage.**

Inserting expression(9) into equation(2), one deduces $G_{2D}(x,k_y,z_0)$ as a convolution of $\partial_z E_{2D}(x,-L_{//}/2,z_0)$ with a Green's function for the diffusion equation [Colas2012]:

$$G_{2D}(x, k_y, z_0) = \frac{\varepsilon_{//}}{\varepsilon_\perp} \int_0^{L_\perp} \partial_z E_{2D}(x', -L_{//}/2, z_0) \frac{\sinh(k_y x_{\min}) \sinh(k_y (L_\perp - x_{\max}))}{k_y \sinh(k_y L_\perp)} dx'$$

$$(19)$$

Where $x_{min} = \min(x,x')$ and $x_{max} = \max(x,x')$.

For the ASDEX-Upgrade parametersin [Křivská2015], Figures 3plot $G_{2D}$ versus $x$ for two values of $k_y$ and various parallel distances $\delta z_0$ between the emission point and the left wall. The boundary conditions in equation(2) impose $G_{2D}(0, k_y, z_0) = 0$ and $G_{2D}(L_\perp, k_y, z_0) = 0$. Between these two radial extremities $G_{2D}$ at fixed $z_0$ exhibits a radial maximum, whose position shifts radially inwards with increasing $\delta z_0$.

Figures 3 show that for fixed $x$ $G_{2D}$ decreases with increasing $\delta z_0$. This is a first evidence of parallel proximity effects in realistic tokamak conditions. SW evanescence ensures that this result is quite general: indeed $\partial_z E_{2D}(x,z=-L_{//}/2,z_0)$ is a decreasing function of $z_0$. From (19) one deduces that this is also the case for $G_{2D}$. When the source point moves towards the right wall formula (13) yields

$$G_{2D}(x, k_y, +L_{//}/2) = 0 \quad (20)$$

The lower curves on Figure 3 reflect this trend. When the source point gets close to the left wall the limit behavior is deduced from formula (17)

$$G_{2D}(x, k_y, -L_{//}/2) \approx 2\int_0^{L_\perp} \partial_{x'} \delta(x') \frac{\sinh(k_y x_{\min})}{k_y} \frac{\sinh(k_y(L_\perp - x_{\max}))}{\sinh(k_y L_\perp)} dx'$$
$$= 2 \frac{\sinh(k_y(L_\perp - x))}{\sinh(k_y L_\perp)} \int_0^{L_\perp} \delta(x') \cosh(k_y x') dx' = \frac{\sinh(k_y(L_\perp - x))}{\sinh(k_y L_\perp)} \quad (21)$$

Expression (21) corresponds to the dashed lines on figures 3.

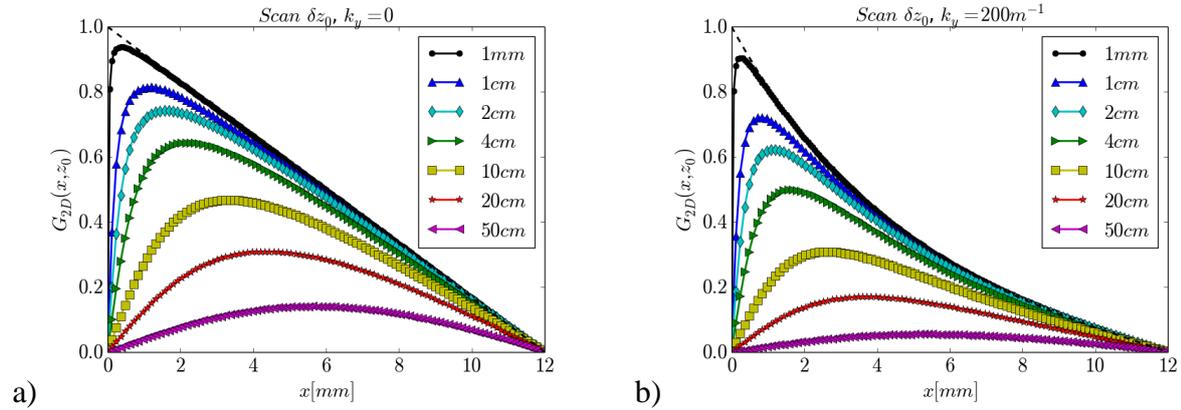

**FIGURE 3.** Green's function $G_{2D}(x,k_y,z_0)$ *versus* radial coordinate $x$ for increasing parallel distance $\delta z_0=(L_{//}/2+z_0)$ from wave emission point $z=z_0$ to left parallel boundary $z=-L_{//}/2$. $x$ is 0 at aperture and increases towards leading edge of antenna limiter at $x=L_\perp=12$mm. Simulation with ASDEX-Upgrade parameters used in [Křivská2015] and (a) $k_y=0$, (b) $k_y=200$m$^{-1}$. Dashed lines: asymptotic expression (21).

From the two limit expressions (20) and (21), we deduce that $0 \leq G_{2D}(x,z_0) \leq \sinh(k_y(L_{//}-x))/\sinh(k_y L_{//}) \leq 1$: $G_{2D}$ is a real positive attenuation factor and $|V_{RF}(x,k_y)| < \int_{-L_{//}/2}^{+L_{//}/2} |E_{//ap}(z,k_y)| dz$.

The way $G_{2D}$ decreases with $\delta z_0$ depends on the input parameters. To quantify these parallel proximity effects, a first indicator is the e-fold parallel decay length $\lambda_z(x)$ of $G_{2D}(x,k_y,z_0)$ at $z_0=0$. In a series of numerical simulations $\lambda_z(x)$ was fitted numerically for 20 values of $x$ from 0 to $L_\perp$. Figure 4 plots $\lambda_z(x)$ averaged over $x$ versus $L_z$, for various parametric scans, exhibiting two regimes. On the low-$L_z$ branch of the curves, if $\delta Z_0 >> X$ while $Z_1 > Z_0$, equation (18) shows that $\partial_z E_{2D}(x,z=-L_{//}/2,z_0)$ decreases as $\sim \exp(-\delta z_0/L_z)$ for all $x$ and so does $G_{2D}(x,k_y,z_0)$. A saturation of $\lambda_z$ is however observed as $L_z$ gets of the order of $L_{//}$. Over the scans of $\varepsilon_\perp$, the saturation level on this opposite branch is found proportional to $L_{//}$. Indeed if $L_{//} << L_z$, $\delta Z_0 << 1$ for all $\delta z_0 < L_{//}$, but all terms matter *a priori* in summation (9). However all the relevant contributions to this summation can be linearized. Expression (20) then ensures that $G_{2D}(x,k_y,z_0)$ decreases linearly as $(1-\delta z_0/L_{//})$

$$G_{2D}(x,k_y,z_0) \approx G_{2D}(x,k_y,z_0=-L_{//}/2)\left(1-\frac{\delta z_0}{L_{//}}\right)$$
$$= \frac{\sinh(k_y(L_\perp-x))}{\sinh(k_y L_\perp)}\left(1-\frac{\delta z_0}{L_{//}}\right) \quad ; \quad L_{//} << L_z \quad (22)$$

Expression (22) shows that in the limit $L_{//} << L_z$ the characteristic length $L_z$ plays no role in the SSWICH-SW problem. From figure 6 and the above estimates, one concludes that $\lambda_z < \min(L_z,L_{//})$.

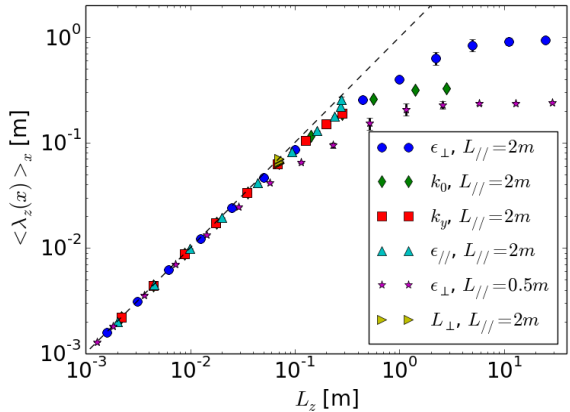

**FIGURE 4.** Parallel e-fold decay length $\lambda_z(x)$ of $G_{2D}(x,k_y,z_0)$ at $z_0=0$ fitted numerically and averaged over 20 values of $x$, versus $L_z$ from eq. (5), for 6 scans of the main parameters in the asymptotic model, each identified by a marker type. Error bars: dispersion of $\lambda_z(x)$ over $x$.

Another quantitative indicator of parallel proximity effects, the parallel gradient length of $G_{2D}$ at $\delta z_0=0$, is plotted on figure 5. Below we seek an upper bound on this gradient length. The parallel gradient of $G_{2D}$ is expressed as

$$\partial_{z_0} G_{2D}(x,k_y,z_0) = \frac{\varepsilon_{//}}{\varepsilon_\perp} \int_0^{L_\perp} \partial^2_{zz_0} E_{2D}(x',-L_{//}/2,z_0) \frac{\sinh(k_y x_{\min})}{k_y} \frac{\sinh(k_y(L_\perp-x_{\max}))}{\sinh(k_y L_\perp)} dx' \quad (23)$$

Where $\partial^2_{zz_0} E(x,-L_{//}/2,z_0)$ is built from

$$\partial_{ZZ}F_{2D}(X,Z) = -\frac{X}{\pi R^3}\left[RK_2(R) - Z^2 K_3(R)\right] \tag{24}$$

From (23) and the above analysis one deduces that for $L_z \ll L_{//}$, $\partial_{z0}G_{2D}$ scales as $L_z^{-1}$ when all other parameters are kept constant, while for $L_z \gg L_{//}$

$$\partial_{z_0}G_{2D}(x,k_y,z_0) \approx -G_{2D}(x,k_y,z_0 = -L_{//}/2)/L_{//} \quad ; \quad L_{//} \ll L_z \tag{25}$$

From figure 3 one also anticipates very steep gradients as $x$ gets very small. Concretely, this means that minimizing $|V_{RF}|$ near $x=0$ gets equivalent to cancelling $E_{//ap}$ at the parallel extremities of the input RF field map, consistent with the optimization criterion proposed in [Bobkov2015]. One can show that an upper bound for $G_{2D}$ gradient length is given by

$$L_{z\max} = \min\left[\frac{\pi x}{2L_x}L_z, L_z / I\left(\frac{L_\perp}{L_x}, k_y L_x\right), L_{//}\right]$$
$$I\left(\frac{L_\perp}{L_x}, k_y L_x\right) = 2\int_0^{L_\perp/L_x} \partial_{ZZ}^2 F_{2D}(X, Z=0)\frac{\sinh(k_y L_x X)}{k_y L_x}dX \tag{26}$$

Figure 5 illustrates numerically this upper bound over four orders of magnitude, for various scans of the main parameters in the SSWICH asymptotic model.

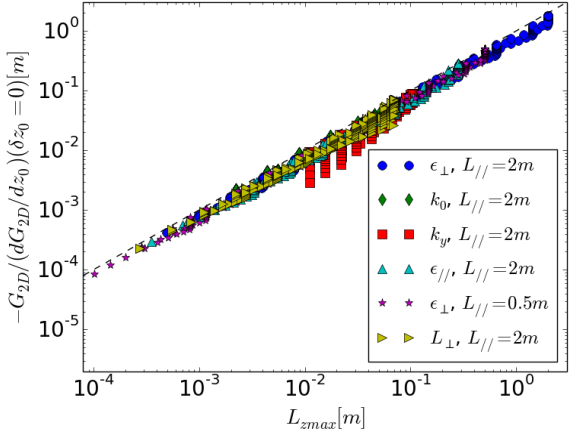

**FIGURE 5.** Parallel gradient length of $G_{2D}(x,k_y,z_0)$ fitted numerically at $\delta z_0=0$, *versus* upper bound $L_{zmax}$ from eq. (26). For each simulation, 19 points are plotted, for $x$ values located every 5% of $L_\perp$. Marker types indicate simulation series with one parameter scanned.

### 3.3 Example: two-peak asymmetric input field map

As a more concrete application, we consider a test case qualitatively similar to the symmetry-breaking experiments in [Colas2013], [Bobkov2015]. We compute $V_{RF}$ using the ASDEX-Upgrade parameters and $k_y=0$, for an input field map composed of two Gaussian peaks: $E_{//ap}(z) = E_1(z) + E_2(z)$ with

$$E_j(z) = E_j \exp\left(-\frac{(z-z_j)^2}{\Delta z^2}\right), j=1, 2 \qquad (27)$$

The parallel half-widths at $1/e$ are chosen as $\Delta z=2$cm for both peaks. The first peak is centered at $z_1=-0.2$m close to the left boundary, while the second is at $z_2=+0.2$m. Initially the field map is toroidally antisymmetric: the two peaks are of opposite signs and equal magnitude, so that $\int E_{//ap}.dl=0$. Namely $V_1=\int E_1(z)dz=-230$V and $V_2=\int E_2(z)dz=+230$V. This initial field map is then progressively unbalanced, by keeping the same shape for the peaks and adding the same voltage $0.5\Delta V$ to $V_1$ and $V_2$, such that $\int E_{//ap}.dl=\Delta V$. Figure 6 plots the resulting $V_{RF}$ at the left boundary versus $x$ for several values of $\Delta V$. Figure 7 shows $V_{RF}$ at both field line extremities *versus* $\Delta V$ for selected $x$. For this series of asymmetric field maps the amplitudes of sheath oscillating voltages are generally different at the two extremities of the same open magnetic field line. Consistent with equation (4), they become equal for $\Delta V=0$, *i.e.* for a toroidally anti-symmetric input field map. As already noticed in [D'Ippolito2006], sheath oscillations exist despite $\int E_{//}.dl$ being null on every open field line. The superposition principle implies that $V_{RF}$ varies linearly with $\Delta V$. The slope of this variation depends on $x$, and $V_{RF}$ evolves in opposite ways at both field line extremities over the same variation of $\int E_{//ap}.dl$. By choosing an appropriate $\Delta V$ it is possible to cancel $V_{RF}$ at given $x$ on the left boundary. For that the two peaks must be of opposite signs and the magnitude of the right peak should be roughly 10 to 20 times that of the left peak, consistent with a parallel proximity effect. The exact peak ratio depends on $x$, so it is not possible to cancel the sheath oscillations everywhere at the same time. For symmetry reasons one should use negative $\Delta V$ to reduce $V_{RF}$ at the right boundary. Therefore, with the considered field maps, it is not possible to mitigate RF-sheath excitation simultaneously at both field line extremities. It is neither possible to cancel completely $|V_{RF}|$ at any place when complex $\Delta V$ is applied, *i.e.* when the two peaks are not in perfect phase opposition.

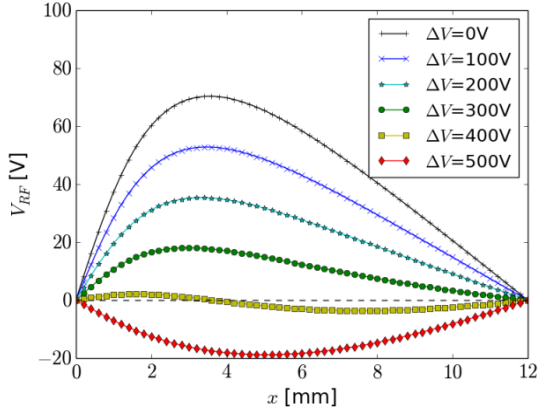

**Figure 6**: Sheath oscillating voltage at the left boundary versus radial distance to antenna aperture. Calculations performed with ASDEX-Upgrade parameters, $k_y$=0 and two-peak input field maps from equation (27). Five curves are showed, for several values of $\Delta V = \int E_{//\text{ap}}.dl$ over the input field map.

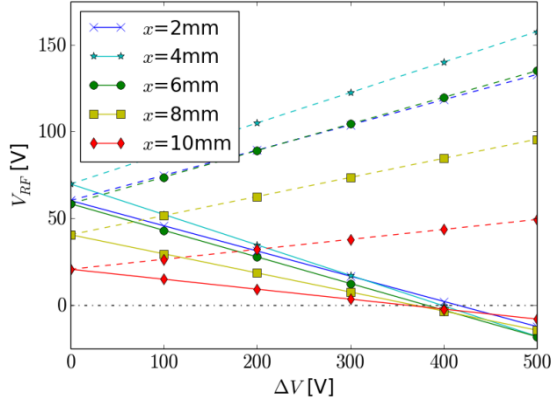

**Figure 7**: Sheath oscillating voltages at selected radial positions $x$ versus $\Delta V = \int E_{//\text{ap}}.dl$ over the two-peak input field map from equation (27). Solid lines: left boundary. Dashed lines: right boundary. Calculations performed with ASDEX-Upgrade parameters and $k_y$=0.

## 4. EXTENSION TO 3 DIMENSIONS

For more realistic description of the RF-sheath excitation, our geometry can be extended to 3D parallelepiped simulation domains. Throughout this part the parallel and radial dimensions $L_{//}$ and $L_\perp$ are the same as in 2D, while the poloidal extent of the domain is infinite. The transverse Laplace operator is redefined as $\Delta_\perp = \partial_{xx}^2 + \partial_{yy}^2$, while both $E_{//}$ and $V_{RF}$ are assumed to vanish for $y \rightarrow \pm\infty$. Equation (3) now consists of a surface integral over a 2D input RF field map $E_{//\text{ap}}(y,z)$

$$V_{RF}(x,y,z=\pm L_{//}/2) = \pm \int_{-\infty}^{+\infty} dy_0 \int_{-L_{//}/2}^{L_{//}/2} E_{//ap}(y_0,z_0) G_{3D}(x, y-y_0, \mp z_0) dz_0 \quad (28)$$

The 3D Green's function $G_{3D}(x,y,z_0)$ has the dimension of a wavevector, and is obtained for the elementary excitation $E_{//\text{ap}}(y,z) = \delta(y)\delta(z-z_0)$. The 3D model exhibits the same characteristic scale-lengths as the 2D model, except that $k_y$=0 is assumed and poloidal coordinate $y$ will appear in the spatial dependences.

## 4.1 Green's function in 3D

The 3D RF field pattern $E_{3D}(x,y,z,z_0)$ is obtained using the same method as in 2D. It is most easily expressed using the normalized quantities $X=x/L_x$, $Y=y/L_x$ and $Z=z/L_z$

$$E_{3D}(x,y,z,z_0) = L_x^{-1} L_z^{-1} \sum_{n=-\infty}^{+\infty} (-1)^n F_{3D}(X,Y,Z-Z_n) \quad ; \quad Z_n = \left[ nL_{//} + (-1)^n z_0 \right]/L_z \tag{29}$$

with

$$F_{3D}(X,Y,Z) = \frac{1}{2\pi} \frac{X(1+R)}{R^3} \exp(-R) \quad ; \quad R = \sqrt{X^2 + Y^2 + Z^2} \tag{30}$$

$F_{3D}$ is null in $X=0$, except in $Y=Z=0$ where it exhibits a singularity. Decay as $\exp(-R)$ is found for large $R$.

$\partial_z E_{3D}(x,y,z,z_0)$ is computed using

$$\partial_Z F_{3D}(X,Y,Z) = -\frac{XZ(3+3R+R^2)}{2\pi R^5} \exp(-R) \tag{31}$$

whence

$$G_{3D}(x,y,z_0) = \frac{\varepsilon_{//}}{\varepsilon_\perp} \int_{-\infty}^{+\infty} \int_0^{L_\perp} \partial_z E_{3D}(x',y',z=-L_{//}/2,z_0) H(x,x',y-y') dx' dy' \tag{32}$$

With the 2D solution of equation (2) given by [Durand1966] p.265

$$H(x,x',y) = \operatorname{argtanh}\left[ \frac{\sin(\pi x/L_\perp) \sin(\pi x'/L_\perp)}{\cosh(\pi y/L_\perp) - \cos(\pi x/L_\perp)\cos(\pi x'/L_\perp)} \right] \tag{33}$$

Figures 8 map $G_{3D}(x,y,z_0)$ *versus* $(x,y)$ as obtained numerically for the ASDEX-Upgrade simulation parameters and three values of $\delta z_0$. At given $(x,z_0)$ $G_{3D}$ is a decreasing function of the poloidal distance $|y|$ from wave emission point to observation point. Over this scan $V_{RF}$ at a given altitude involves the $E_{//\mathrm{ap}}$ values within less than 1.5cm from this altitude. Figures 8 also illustrate how $G_{3D}$ decreases in magnitude, expands in the poloidal direction while its radial maximum moves away from the aperture with increasing parallel distance from wave-emitting point to sheath wall. Let us now quantify these properties.

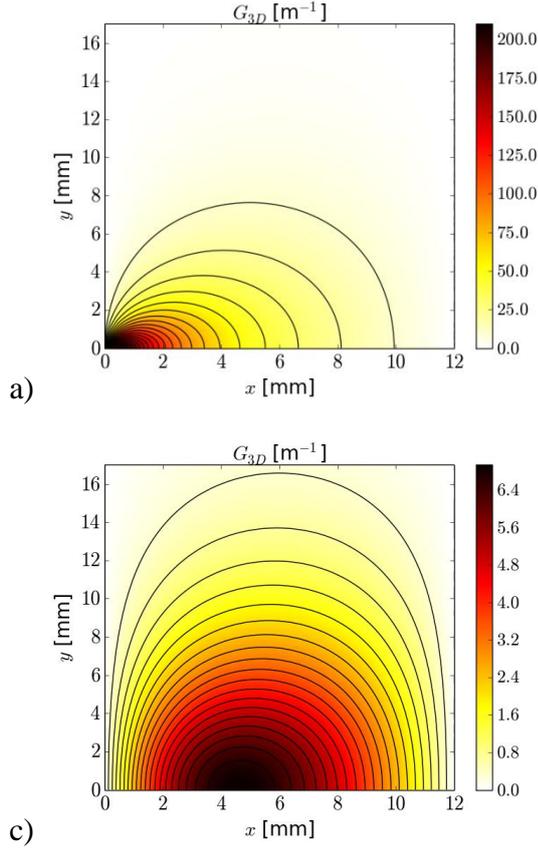

**FIGURE 8.** 3D Green's function $G_{3D}(x,y,z_0)$ *versus* transverse coordinates $(x,y)$, as evaluated numerically using ASDEX-Upgrade simulation parameters in [Křivská2015] and parallel distances (a) $\delta z_0=(L_{//}/2+z_0)=2.5$cm; (b) $\delta z_0=10$cm and (c) $\delta z_0=33$cm. Contour lines are located every 5% of the maximum value over the map.

### 4.2 Evolution with $\delta z_0$

SW evanescence ensures that $G_{3D}$ decreases with $\delta z_0$ at fixed $(x,y)$. Since $\delta(y) = \frac{1}{2\pi}\int_{-\infty}^{+\infty}\exp(ik_y y)dk_y$ the 2D and 3D Green's functions are Fourier transforms of each other

$$G_{3D}(x, y, z_0) = \frac{1}{2\pi}\int_{-\infty}^{+\infty} G_{2D}(x, k_y, z_0)\exp(i k_y y)dk_y \qquad (34)$$

From our 2D analysis in part 3, one deduces that $G_{3D}$ is null for $z_0=+L_{//}/2$. In the opposite limit $z_0\to -L_{//}/2$, one gets from relation(21) ([Gradshteyn1980] p.504)

$$\begin{aligned}G_{3D}(x, y, -L_{//}/2) &\approx \frac{1}{2\pi}\int_{-\infty}^{+\infty}\frac{\sinh(k_y(L_\perp - x))}{\sinh(k_y L_\perp)}\exp(ik_y y)dk_y \\ &= \frac{1}{2L_\perp}\frac{\sin(\pi x/L_\perp)}{\cosh(\pi y/L_\perp)-\cos(\pi x/L_\perp)}\end{aligned} \qquad (35)$$

And if $L_z \gg L_{//}$ while $Y^2 \ll 1$, one anticipates a linear decay with parallel distance $\delta z_0$.

$$G_{3D}(x,y,z_0) \approx \frac{1-\delta z_0/L_{//}}{2L_\perp} \frac{\sin(\pi x/L_\perp)}{\cosh(\pi y/L_\perp)-\cos(\pi x/L_\perp)} \qquad (36)$$

Figure 9 illustrates the limit expression of $G_{3D}(x,y,-L_{//}/2)$ in (34). No wave evanescence is involved: $L_x$ and $L_z$ disappear from the problem, all coordinates can be normalized by the only remaining characteristic length $L_\perp$. Since the emission point is infinitely close to the left wall, $G_{3D}$ exhibits a singularity in $(x,y)=(0,0)$. Concretely this means that minimizing $|V_{RF}|$ near $x=0$ is equivalent to cancelling the local $E_{//ap}$ near the sheath observation point in the input field map.

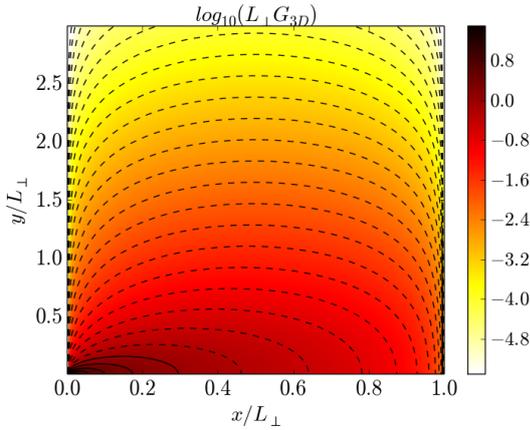

**FIGURE 9.** 2D (radial,poloidal) map of $L_\perp G_{3D}(x,y,-L_{//}/2)$ from formula (35) in logarithmic scale, versus normalized coordinates $(x/L_\perp, y/L_\perp)$. Contour lines: between two consecutive curves the function decreases by a factor $10^{1/4} \approx 1.78$. Solid lines correspond to $L_\perp G_{3D}>1$, dashed lines to $L_\perp G_{3D}<1$.

Poloidal integration of $G_{3D}$ yields

$$\int_{-\infty}^{+\infty} G_{3D}(x,y,z_0)dy = G_{2D}(x,k_y=0,z_0) \qquad (37)$$

From the 2D analysis, one deduces that for $\delta z_0/L_z >> (x^2+y^2)^{1/2}/L_x$ and $L_{//}/2-z_0>>L_z^2$, the poloidal integral of $G_{3D}$ decays as $\exp(-\delta z_0/L_z)$ for large $\delta z_0$. The upper bound $L_{zmax}$ from expression (26) is also valid.

### 4.3 Poloidal decay lengths, relevance of 2D simulations

Surface integral (28) can be seen as a weighted sum of line integrals over several open magnetic flux tubes instead of one in the previous approaches. It is worth estimating how many of these open magnetic field lines do really matter in expression(28). A related issue is the validity of 2D SSWICH-SW simulations in part 3 in comparison with the more accurate, but more computationally demanding 3D simulations in part 4. This amounts to evaluating the poloidal extent of $G_{3D}$ at fixed $(x,z_0)$.

For $\delta z_0=0$ formula(35) features a minimal poloidal extent of $G_{3D}$ in the absence of SW evanescence. The half-width at 1/e can be evaluated analytically as

$$\frac{\Delta y}{L_\perp} = \frac{1}{\pi} a\cosh\left[e + (1-e)\cos\left(\frac{\pi x}{L_\perp}\right)\right] \qquad (38)$$

Expression(38) shows that $\Delta y < 0.7 L_\perp$ over the whole radial range of the simulation domain.

As the wave emission point moves away from the sheath wall, SW evanescence broadens $G_{3D}$ in the poloidal direction. Formula(32) presents $G_{3D}(x,y,z_0)$ as the convolution of $\partial_z E_{3D}(x',y,z=-L_{//}/2,z_0)$ with $H(x,x',y)$. $\partial_z E_{3D}$ scales as $\sim R^{-5}$ for small $R$ and as $\sim \exp(-R)$ for large $R$. An upper bound for its poloidal extent is therefore

$$L_{E\max}(x',\delta z_0) = \min\left[0.7 L_x \sqrt{\left(\frac{\delta z_0}{L_z}\right)^2 + \left(\frac{x'}{L_x}\right)^2},\, L_x\sqrt{1+2\sqrt{\left(\frac{\delta z_0}{L_z}\right)^2 + \left(\frac{x'}{L_x}\right)^2}}\right] \qquad (39)$$

The poloidal half-width of $H$ can be expressed explicitly as

$$L_H(x,x') = \operatorname{arccosh}\left[\frac{\sin(\pi x/L_\perp)\sin(\pi x'/L_\perp)}{\tanh[H(x,x',0)/e]} + \cos\left(\frac{\pi x}{L_\perp}\right)\cos\left(\frac{\pi x'}{L_\perp}\right)\right] \qquad (40)$$

$L_H$ is an increasing function of $|x-x'|/L_\perp$. The source term for $G_{3D}$ at point $x$ is present from $x'=0$ to $x'=\min(x+L_x, L_\perp)$ (pessimistic estimate). One can then put an upper bound on the half-poloidal with for $G_{3D}$

$$\Delta y < \max[L_H(x,0), L_H(x, \min(x+L_x, L_\perp)), L_{E\max}(\min(x+L_x, L_\perp), \delta z_0)] \qquad (41)$$

The above estimates are assessed numerically on figure 10. In this exercise 2D (radial, poloidal) maps of $G_{3D}$ at constant $z_0$ were simulated numerically over several scans of the main simulation parameters. From each map $\Delta y$ was fitted at several radial positions. Over the tested parametric domain inequality (41) is well verified, and the upper bound is sometimes pessimistic by a factor 2 or 3.

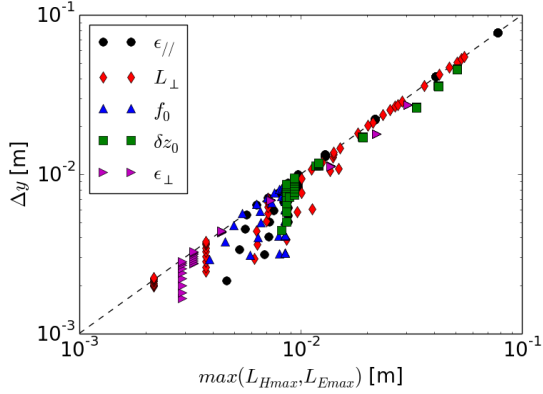

**FIGURE 10.** Half poloidal width $\Delta y$ at $1/e$, fitted numerically from simulated 2D (radial, poloidal) maps for $G_{3D}$. For each simulation $\Delta y$ was fitted at 9 radial positions ranging from $x/L_\perp=0.1$ to 0.9 and plotted versus $\max(L_{Hmax}, L_{Emax})$ from formula (41). Each series of points refers to a scan of one simulation parameter indicated in the legend.

For poloidal structures larger than $\max(L_{Hmax}, L_{Emax})$ in the input field map, $G_{3D}$ can be considerably simplified using (34).

$$G_{3D}(x,y,z_0) \approx G_{2D}(x,k_y=0,z_0)\delta(y) \quad (42)$$

Surface integral (28) then reduces to a weighted integral along one single field line, located at the same altitude as the observation point. Consequently above a critical length, the poloidal structures of $V_{RF}$ reflect those of $E_{//\mathrm{ap}}$ near the parallel extremities of the input field map. Smaller scales below the critical length in the input RF field map are smoothed and contribute less to $V_{RF}$.

## 5. DISCUSSION AND CONCLUSION

### 5.1 Practical implications

Within the existing asymptotic SSWICH-SW model recalled in part 2, RF oscillations $V_{RF}$ of the sheath voltage at any open field line extremity can be re-expressed as a sum of individual contributions by each emitting point in the parallel RF electric field map $E_{//\mathrm{ap}}(\mathbf{r_0})$ radiated by an IC antenna. This re-formulation offers a simple alternative to the "double probe" criterion $\tilde{V}=|\int E_{//}.dl|$ for assessing sheath RF voltages closer to the first principles in the "wide-sheath" limit. For practical $V_{RF}$ computations with realistic input field maps, this alternative method is generally less efficient numerically than the Fourier technique in [Myra2010], [Colas2012] or the Finite Element Method in [Kohno2012], [Jacquot2014]. It allows however to reveal and quantify spatial proximity effects in the excitation of oscillating sheath voltages. Indeed, for the first time to our knowledge, proposed formula(3) consists of a *weighted* integral of $E_{//\mathrm{ap}}$: Slow Wave (SW) evanescence causes point-source contributions (or Green's functions for $V_{RF}$) to decrease with increasing parallel and poloidal distances from wave emission point to sheath walls. As a test case in a parallelepiped box filled with homogeneous cold magnetized plasma, 2D and 3D Green's functions were determined explicitly in the limit of an emission point very close to the sheath walls, and their spatial variations were quantified numerically as a function of characteristic lengths in our model.

Poloidal decay lengths for $V_{RF}$ involve the radial protrusion $L_\perp$ of antenna side limiters, as well as the transverse SW evanescence length $L_x$, with extra broadening due to the parallel evanescence. In realistic situations, these poloidal decay lengths are much lower than the typical vertical extent of ICRF antennas, e.g. less than 1.5cm for our ASDEX-Upgrade example. This is qualitatively consistent with experimental observations that RF-induced SOL modifications are mainly observed on magnetic field lines passing in front of the antenna box, while they are absent on field lines connecting above or below the box aperture [Jacquot2014], [Cziegler2012], [Kubič2013]. If poloidal structures in the input field map are larger than the decay length, independent 2DSSWICH-SW simulations at each altitude fairly approximate the full 3D models, while the 2D input RF field map retains 3D information about the global antenna geometry.

The parallel decay lengths for $V_{RF}$ mainly involve the minimum between the connection length $L_{//}$ and the parallel SW evanescence length $L_z$. This result is not specific of SSWICH-SW: the role of $L_z$ is probably generic of any model featuring SW evanescence. Within the "wide sheath limit", this generalizes to any parallel distribution of $E_{//\text{ap}}$ the role of the ion skin depth pointed out in [Myra2010]. $L_z$ is related to the transverse coupling of adjacent open magnetic field lines via $\varepsilon_\perp$ in equation 1. Such transverse coupling was absent in the simplest "double probe" model. In SW propagation, decoupling is only obtained at the LH resonance ($\varepsilon_\perp=0$) and leads to infinite $L_z$. This paper also evidenced other parametric dependences of the parallel decay lengths, e.g. with the radial distance to the aperture and the radial extension of the lateral walls. Typical parallel decay lengths are always smaller than typical antenna parallel extensions. Consequently, when the radiated $E_{//ap}$ map exhibits parallel anti-symmetry, an attenuation factor prevents the cancellation of the relevant integral for $V_{RF}$ in equation(4). Sheath oscillations therefore persist, while the previous formula predicts that $\int E_{//}.dl=0$. Similar cases were already evidenced in [D'Ippolito2006]. Besides, the sheaths at the two ends of the same open field line can oscillate differently, depending on the parallel symmetry of $E_{//\text{ap}}$ map. $V_{RF}$ at an IC antenna side limiter appears mainly sensitive to $E_{//ap}$ emission by active or passive conducting elements near this limiter, as experimental observations suggest in [Colas2013][Bobkov2015]. For the realistic simulations of ASDEX-Upgrade in [Křivská2015], a correlation was found between $V_{RF}$ at antenna side limiters and RF field amplitudes at the same altitude, averaged over ~10cm from the side limiters along the parallel direction, whereas the antenna toroidal extension was 66cm. This correlation was independent of the altitude, of the antenna type and of the electrical settings, and mainly depended on the plasma parameters. Toroidal proximity effects could therefore justify current attempts at reducing the local RF fields induced near antenna boxes to attenuate the sheath oscillations in their vicinity [Bobkov2015][Bobkov2016]. The proposed heuristic procedure does fully coincide with our $V_{RF}$-cancellation rule only for very small radial distances $x$ to the input field map. In both cases the optimal setting requires higher RF voltage on the remote straps than on the close ones phased [0,π]. Since one cannot cancel $V_{RF}$ everywhere on the antenna structure, one should carefully choose the spatial locations where to optimize RF-sheaths. Experiments in [Colas2013] and [Bobkov2015] and figure 7 in this paper showed that with two straps (two peaks in our test case), improving the situation at one parallel side of the antenna box likely degrades the situation on the opposite side. Using a 3-strap antenna somehow removes this constraint [Bobkov2016].

In addition to the antenna geometry and its electrical settings, the $V_{RF}$-cancellation rule also involves the local plasma near the antennas: both $L_x$ and $L_z$ decrease with increasing local density near the antenna, through the dielectric constants $\varepsilon_{//}$ and $\varepsilon_{\perp}$. Therefore replacing the plasma by a vacuum layer thicker than $L_x$ in the radial direction could modify the optimal settings. This sensitivity, observed numerically in [Colas2005][Milanesio2013][Colas2014][Lu2016a] [Jacquot2015], is a challenge for quantitative RF-sheath evaluations. RF-sheath optimization may be sensitive to intermittent local density fluctuations naturally present in the tokamak SOL.

Although the above conclusions were reached in a parallelepiped box filled with homogeneous plasma in the "wide sheath" limit, we believe that they persist qualitatively with more complex geometry, density gradients and finite sheath widths. Although Green's functions are harder to determine in these more realistic situations, they still exist in any geometry and in presence of prescribed sheath widths, as long as the physics model remains linear. Therefore Green's functions could be defined for other models in the existing literature on RF sheaths. For Tore Supra, the fully-coupled simulation results with self-consistent sheath widths in [Jacquot2014] were found close in magnitude and spatial structure to the asymptotic first guess provided by the wide sheath approximation. Beyond the "wide-sheath approximation" some parallel proximity effects seem to persist in self-consistent calculations using symmetric Gaussian field maps [Myra2010].

## 5.2 Physical limitations and prospects

The SSWICH-SW model predicts that the direct excitation of sheath oscillations by the evanescent SW is only intense in the ICantenna near RF field [Jacquot2014][Colas2014]and loses efficiencybeyond a parallel distance smaller than $L_z$ from the radiating elements. The experiments in the introduction involved private limiters in this near field. However, RF-induced SOL modifications haveoften been observedexperimentally at parallel distances far larger than $L_z$[Colas2013], [Bobkov2015], [Cziegler2012], [Klepper2013], [Kubič2013], [Lau2013], [Ochoukov2013].To interpret these measurements, extraphysical mechanisms not discussed in the present paper need to be considered.

In very tenuous SOLs below the lower hybrid resonance, the SW becomes propagative [Lu2016a]and can possibly excite RF sheaths at large parallel distances [Myra2008]. Propagative SW can be handled using the Green's function formalism introduced in this paper. However instead of decreasing monotonically with parallel and poloidal distances, the Green's functions may ratheroscillate in a complex way.

At higher densities, the Fast Wave(FW) becomes propagative. It can excite so-called "far-field RF-sheaths" if **B₀** is not strictly normal to the walls[D'Ippolito2008][Kohno2015].The FW can also be incorporated into a generalized Green's function formalism in the "wide sheaths" asymptotic limit. For that purpose the asymptotic RF-sheath boundary conditions need to be extended to account for all RF field polarizations[D'Ippolito2006]. In addition to $E_{//\text{ap}}$, the input RF field map should also include the radiated poloidal electric field. Each RF field component is expected to generate a specific Green's function.Evanescent FW likely exhibit proximity effects. But eachpolarization will feature specific characteristic decay lengths. FW and SWwill likely be coupled upon reflection onto tilted walls[D'Ippolito2008][Kohno2015]. Extension of the SSWICH code to full-wave RF electric fields and shaped sheath walls in 2D is ongoing [Lu2016b].

While this paper discussed the sheath oscillating voltages $V_{RF}$, the deleterious effects in tokamaks ultimately arise from a local DC biasing of the SOL. The sheath rectification in step 3 of SSWICH is intrinsically non-linear and cannot be described with Green's functions. A transport of DC current is able to couple one sheath with its neighbors and the one at the opposite extremity of the same open field line. In the absence of propagating RF waves, Jacquot's paper[Jacquot2014]showed thatDC current transport can still spread a DC bias to remote areas from the near-field regions where SW direct sheath excitation is efficient.Therefore, in order to significantly reduce the rectified DC plasma potential on a given open field line, one should reduce $|V_{RF}|$ at its two extremities as well as on the neighboring field lines. Reducing $|V_{RF}|$ at only one extremity likely drives the circulation of DC current from the high-$|V_{RF}|$ sheath to the low-$|V_{RF}|$ sheath, with limited effects on the DC plasma potential[Jacquot2011].DC currents have been reported in the SOL in variousself-biasing experiments by sheath rectification near active IC antennas [VanNieuwenhove1992], [Gunn2008], [Bobkov2010].

The non-linearity in step 3 further introduces extra propertiesto the fully coupled problem that are absent in the "wide sheath approximation", *e.g.* the existence of multiple solutions or sheath/plasma resonances [Myra2008], [Myra2010], [D'Ippolito2008]. The role of these extra phenomena in tokamak experiments is still unclear.

A European project, outlined in [Colas2014], is ongoing to include all these extra physical mechanisms intomore realistic models of coupled RF wave propagation and DC plasma biasing. Comparison with plasma measurements [Jacquot2014], [Křivská2015]provedessential for codeassessment. The test of a new 3-strap antenna on ASDEX upgrade[Bobkov2016], the restart of the ITER-like antenna on JET[Durodié2012], the commissioning of new antennas on WEST[Hillairet2015], as well as dedicated test beds[Faudot2015] [Crombé2015]will provide new opportunities to assess the SSWICH model over a large diversity of antenna types and plasma regimes, before it can be used to predict the behavior of future antennas.

**Acknowledgements.**This work has been carried out within the framework of the EUROfusion Consortium and has received funding from the European research and training programme under grant agreement N° 633053. The views and opinions expressed herein do not necessarily reflect those of the European Commission.

**Figures**

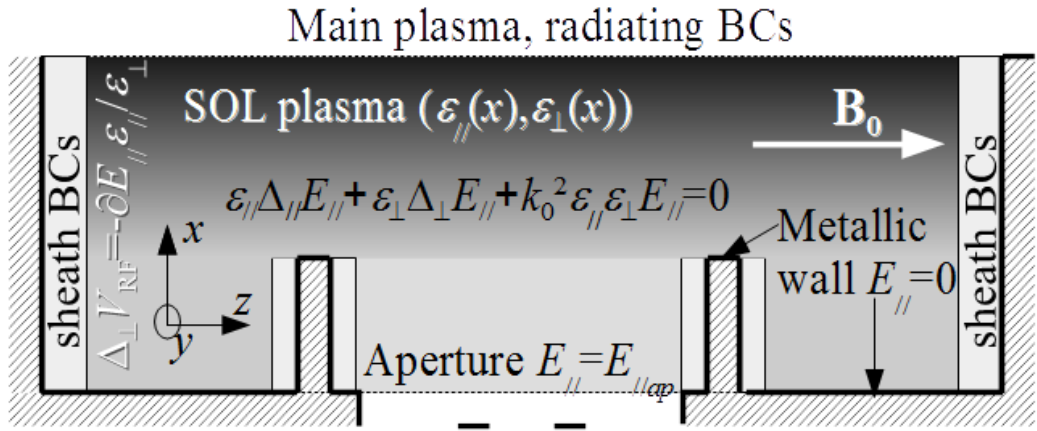

**FIGURE 1**: 2D (radial/parallel) cut into SSWICH general 3D simulation domain (not to scale). Main equations and notations used in the paper. The gray levels are indicative of the local plasma density. Light gray rectangles on boundaries normal to $\mathbf{B_0}$ feature the presence of sheaths, treated as boundary conditions in our formalism.

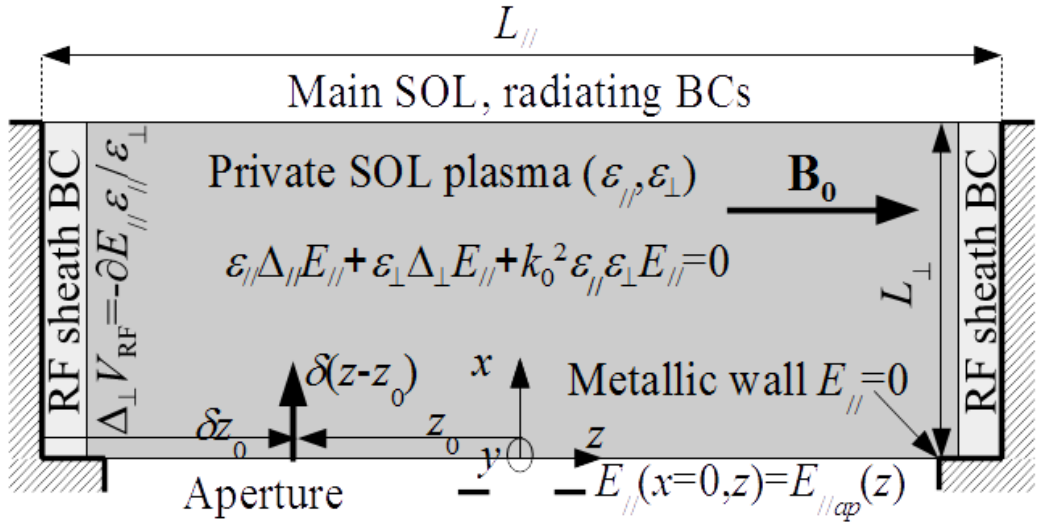

**FIGURE 2**: Generic 2D simulation domain (not to scale). Main equations and notations used in parts 3 and 4. $x=0$ at aperture. Light gray rectangles on boundaries normal to $\mathbf{B_0}$ feature the presence of sheath boundary conditions.

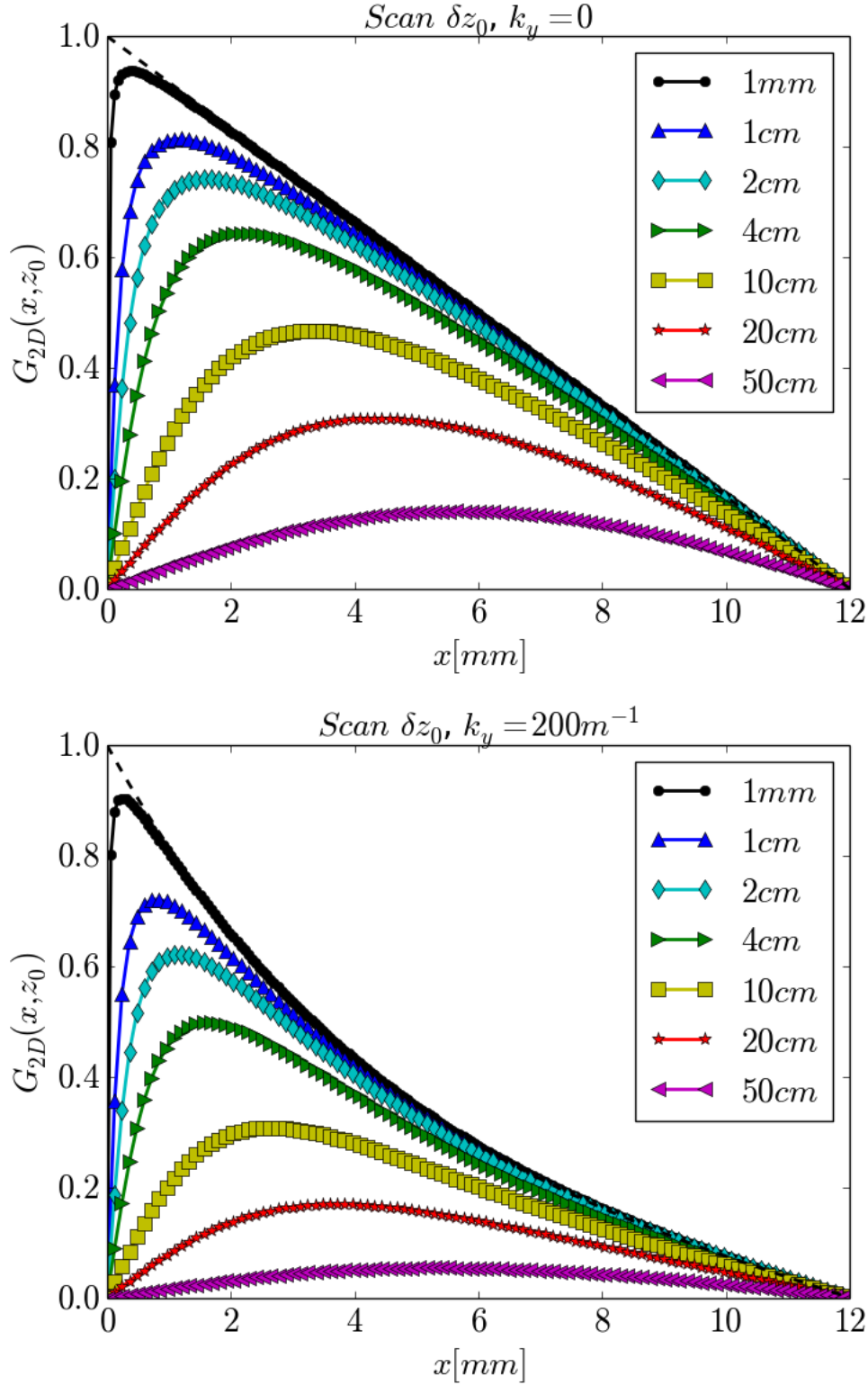

**FIGURE 3.** Green's function $G_{2D}(x,k_y,z_0)$ *versus* radial coordinate $x$ for increasing parallel distance $\delta z_0=(L_{//}/2+z_0)$ from wave emission point $z=z_0$ to left parallel boundary $z=-L_{//}/2$. $x$ is 0 at aperture and increases towards leading edge of antenna limiter at $x=L_\perp=12$mm. Simulation with ASDEX-Upgrade parameters used in [Křivská2015] and (a) $k_y=0$, (b) $k_y=200$m$^{-1}$. Dashed lines: asymptotic expression (21).

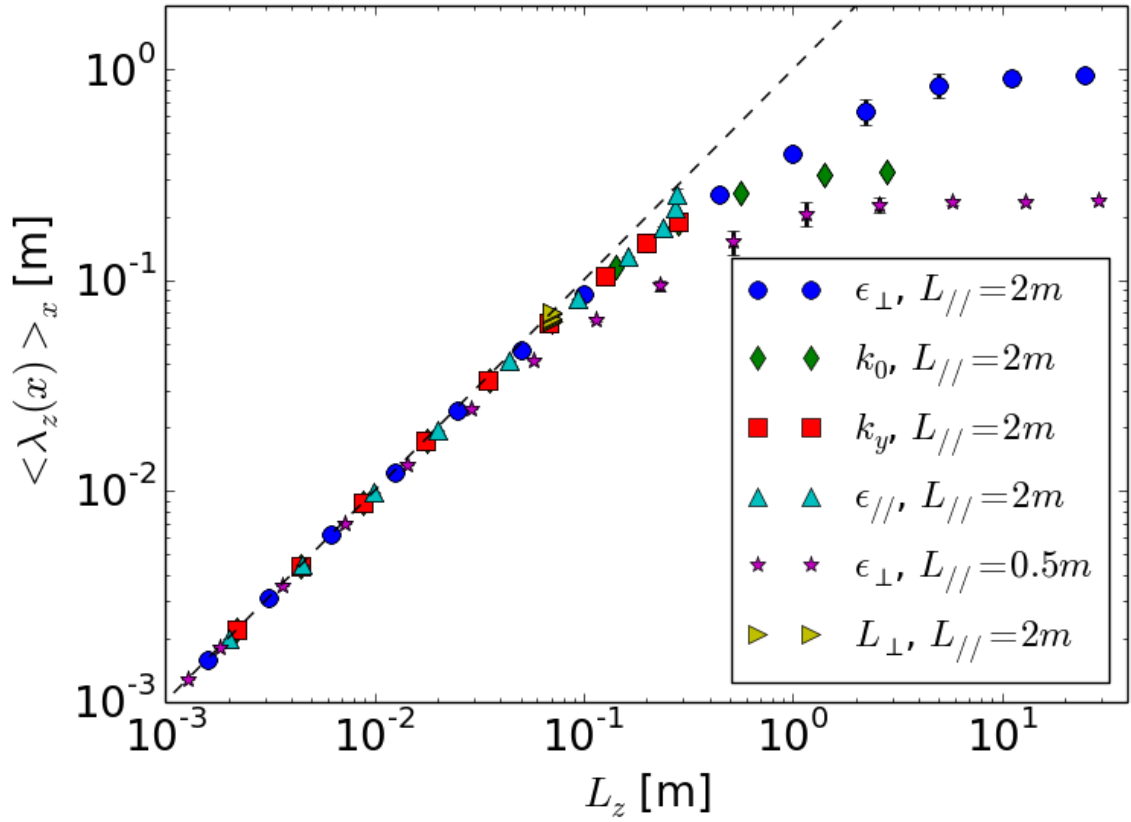

**FIGURE 4.** Parallel e-fold decay length $\lambda_z(x)$ of $G_{2D}(x,k_y,z_0)$ at $z_0=0$ fitted numerically and averaged over 20 values of $x$, versus $L_z$ from eq. (5), for 6 scans of the main parameters in the asymptotic model, each identified by a marker type. Error bars: dispersion of $\lambda_z(x)$ over $x$.

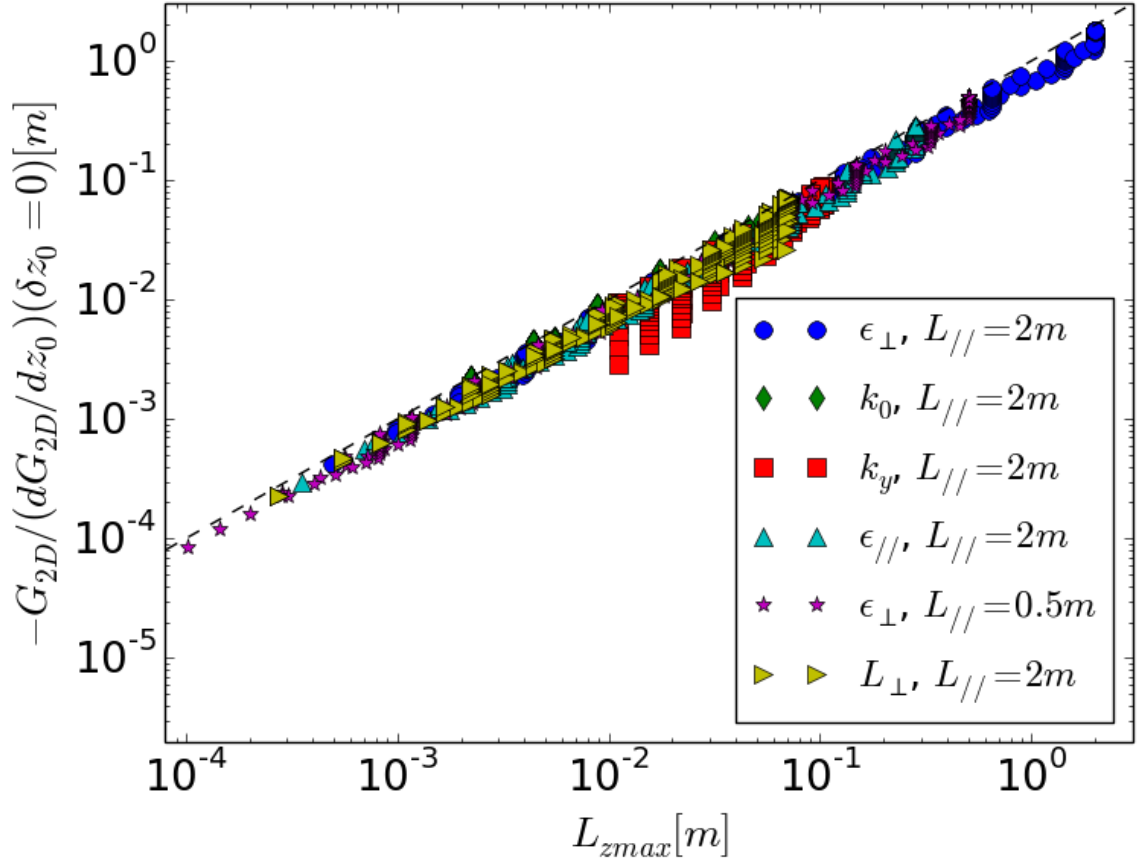

**FIGURE 5.** Parallel gradient length of $G_{2D}(x,k_y,z_0)$ fitted numerically at $\delta z_0=0$, *versus* upper bound $L_{zmax}$ from eq. (26). For each simulation, 19 points are plotted, for $x$ values located every 5% of $L_\perp$. Marker types indicate simulation series with one parameter scanned.

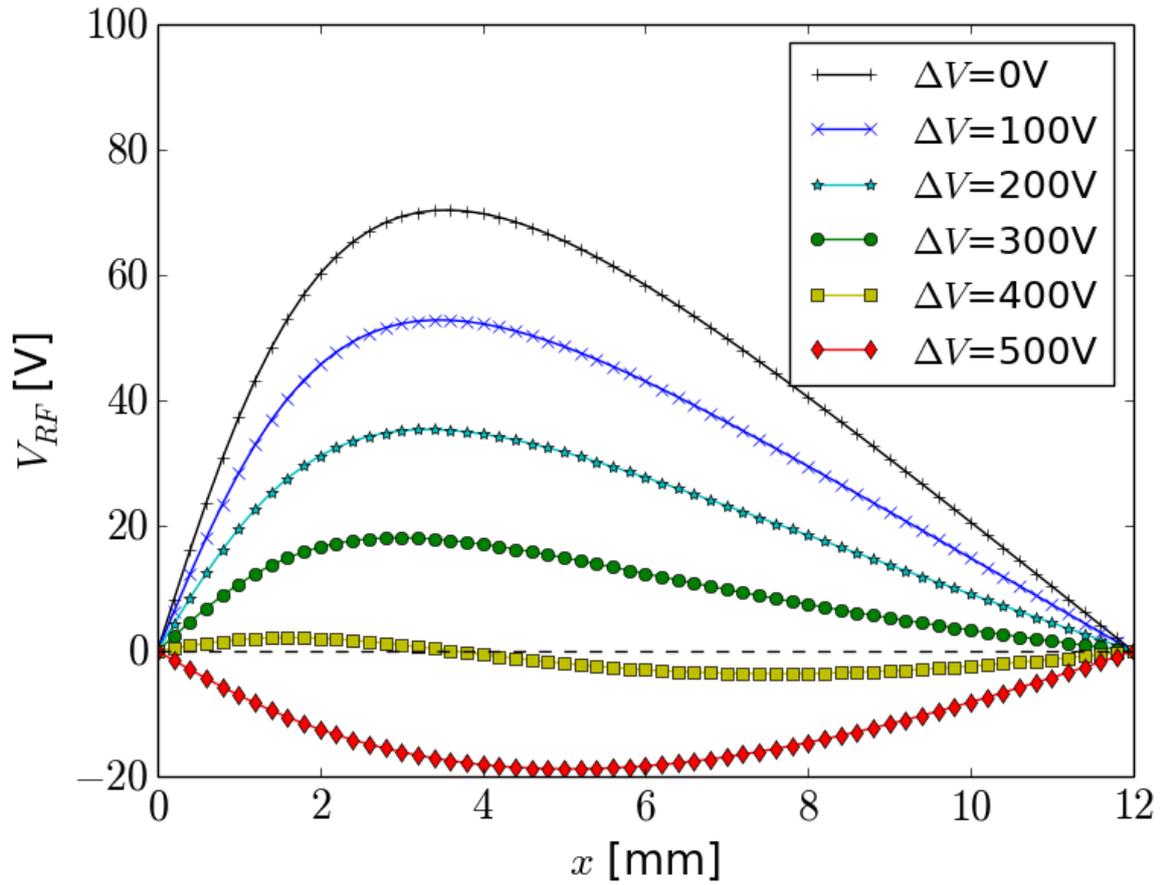

**Figure 6**: Sheath oscillating voltage at the left boundary versus radial distance to antenna aperture. Calculations performed with ASDEX-Upgrade parameters, $k_y$=0 and two-peak input field maps from equation (27). Five curves are showed, for several values of $\Delta V=\int E_{//\text{ap}}.dl$ over the input field map.

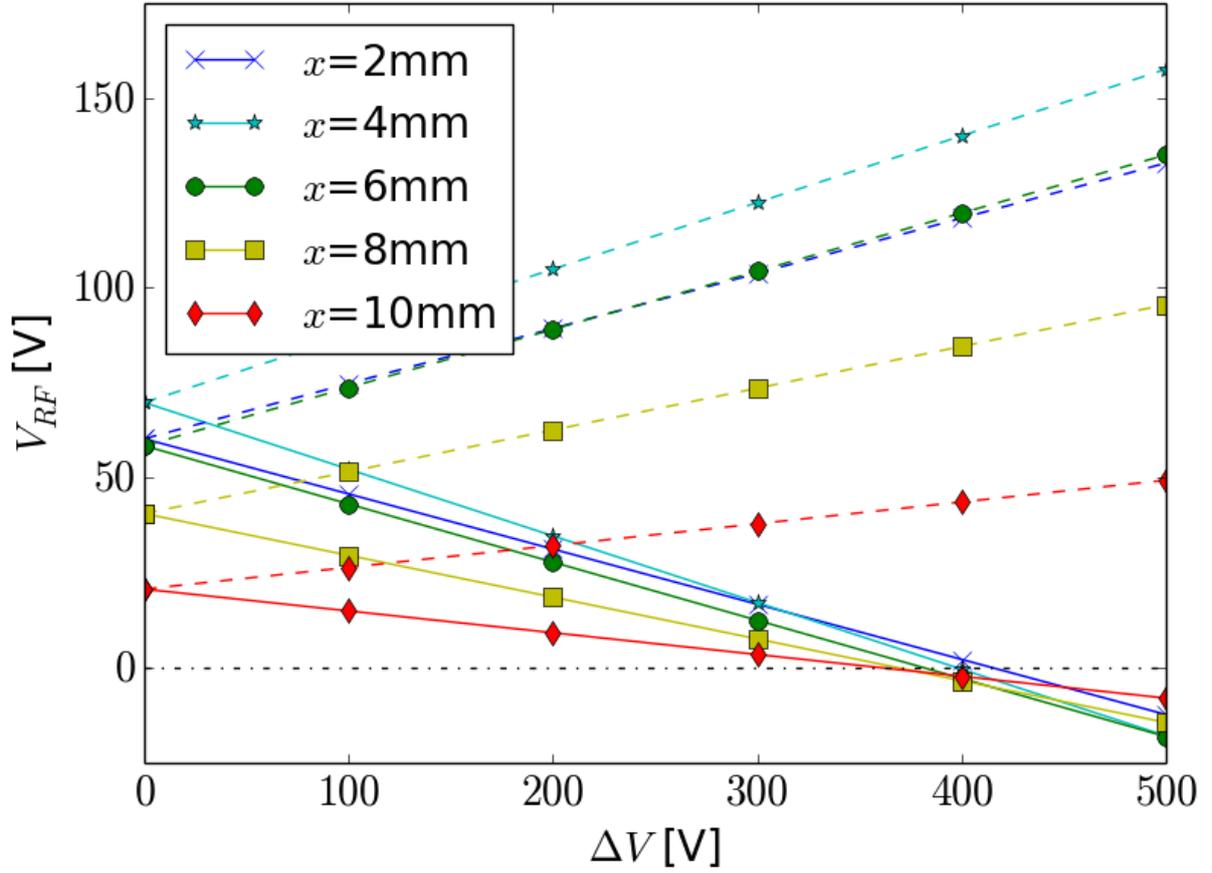

**Figure 7**: Sheath oscillating voltages at selected radial positions *x* versus $\Delta V = \int E_{//ap} \cdot dl$ over the two-peak input field map from equation (27). Solid lines: left boundary. Dashed lines: right boundary. Calculations performed with ASDEX-Upgrade parameters and $k_y=0$.

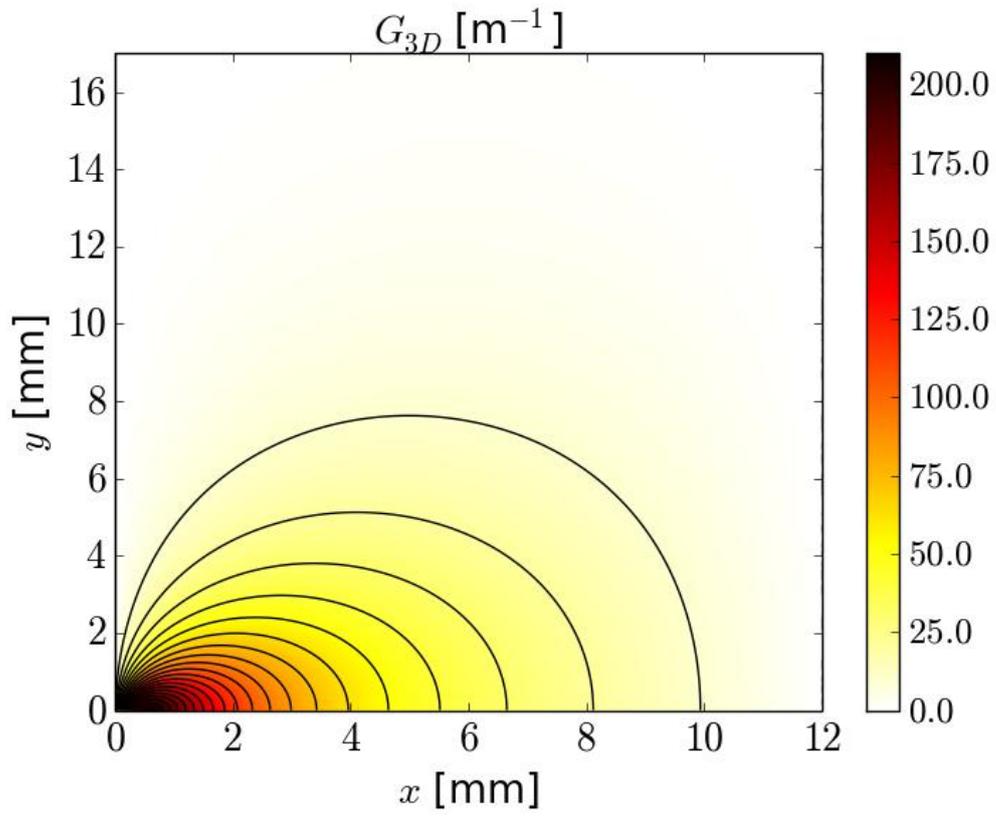

a)

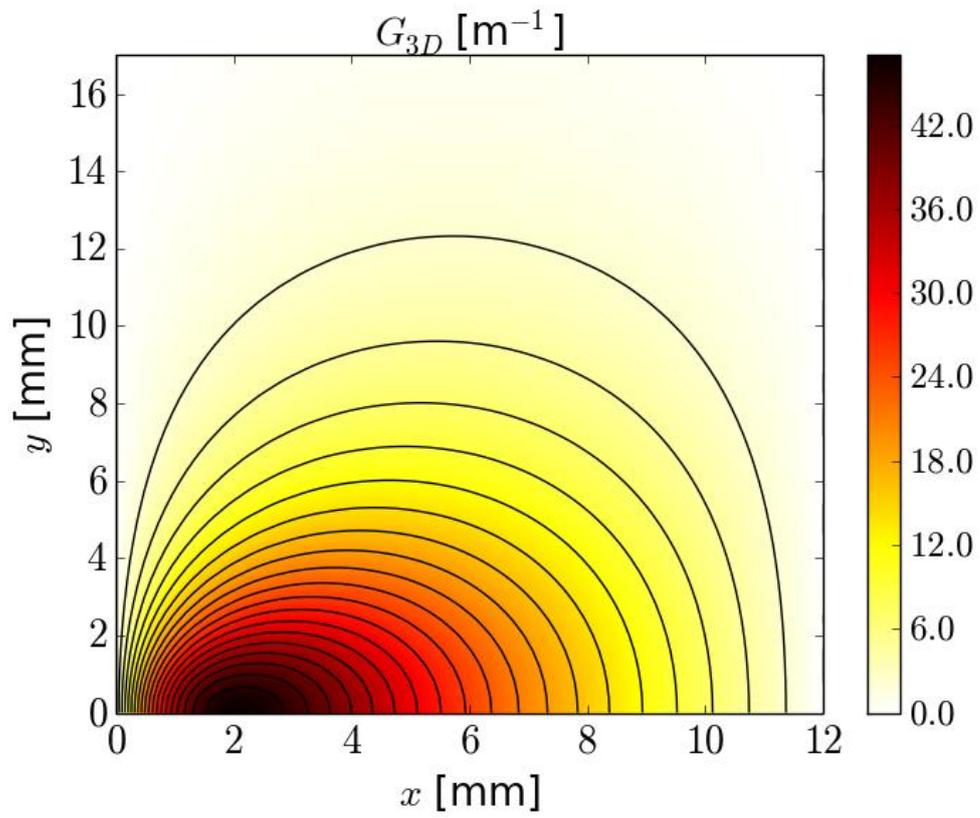

b)

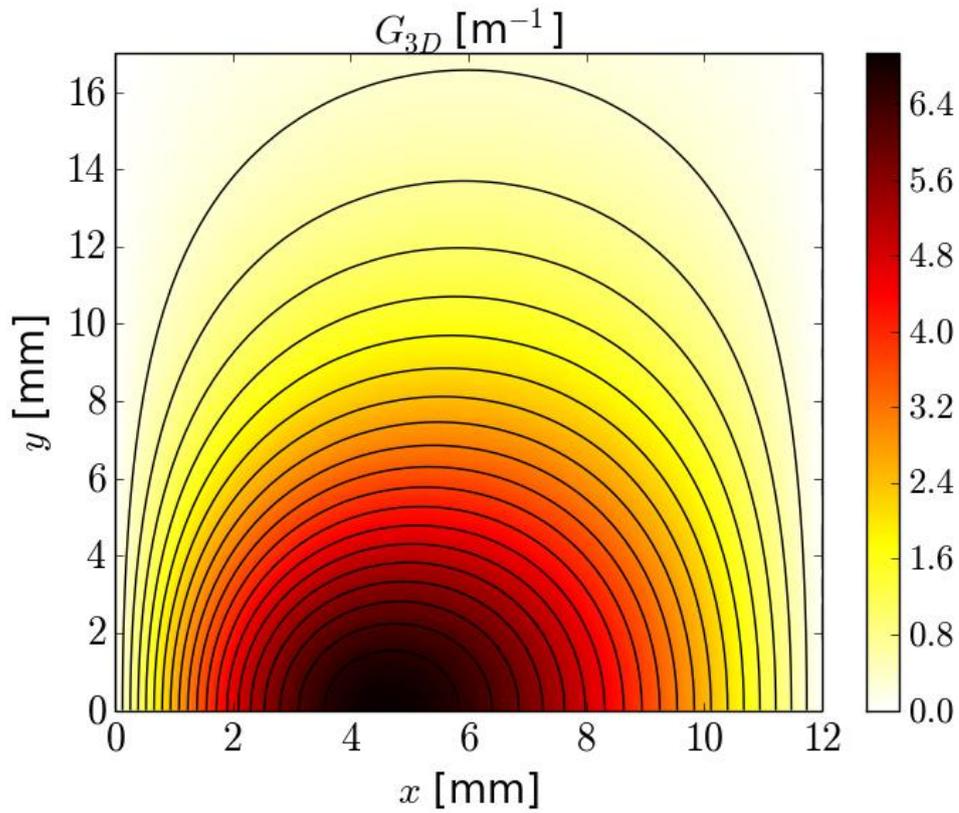

c)

**FIGURE 8.** 3D Green's function $G_{3D}(x,y,z_0)$ *versus* transverse coordinates $(x,y)$, as evaluated numerically using ASDEX-Upgrade simulation parameters in [Křivská2015] and parallel distances (a) $\delta z_0 = (L_{//}/2 + z_0) = 2.5$cm; (b) $\delta z_0 = 10$cm and (c) $\delta z_0 = 33$cm. Contour lines are located every 5% of the maximum value over the map.

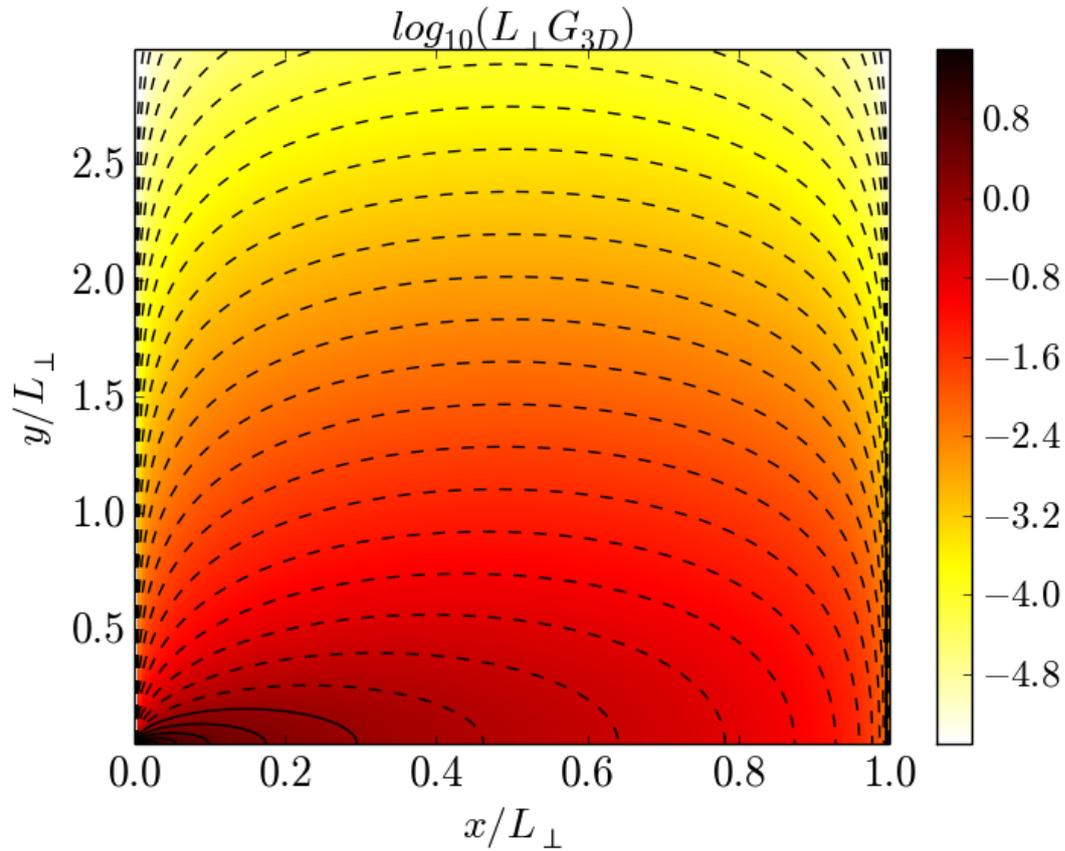

**FIGURE 9.** 2D (radial, poloidal) map of $L_\perp G_{3D}(x,y,-L_{//}/2)$ from formula (35) in logarithmic scale, versus normalized coordinates ($x/L_\perp$, $y/L_\perp$). Contour lines: between two consecutive curves the function decreases by a factor $10^{1/4} \approx 1.78$. Solid lines correspond to $L_\perp G_{3D} > 1$, dashed lines to $L_\perp G_{3D} < 1$.

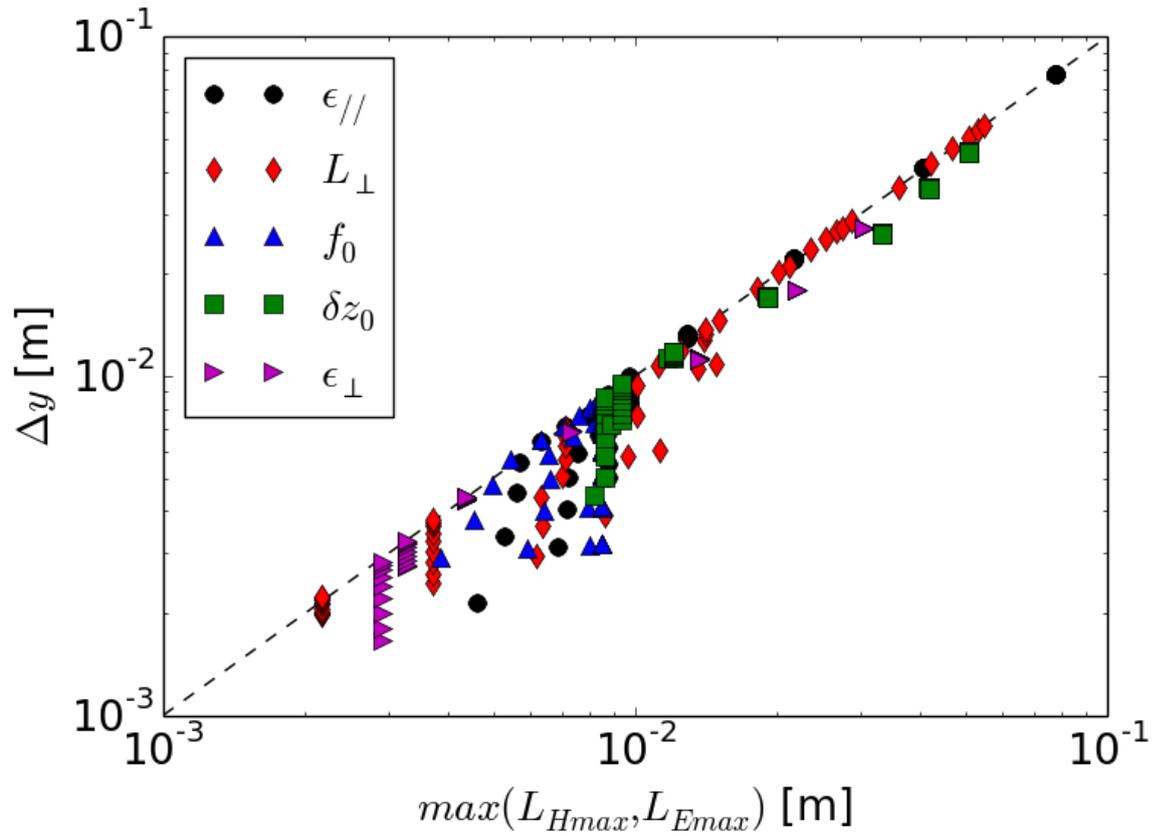

**FIGURE 10.** Half poloidal width $\Delta y$ at $1/e$, fitted numerically from simulated 2D (radial,poloidal) maps for $G_{3D}$. For each simulation $\Delta y$ was fitted at 9 radial positions ranging from $x/L_\perp=0.1$ to $0.9$ and plotted versus $max(L_{Hmax}, L_{Emax})$ from formula (41). Each series of points refers to a scan of one simulation parameter indicated in the legend.